# A Coning Theory of Bullet Motions

James A. Boatright

Revised: March 2018

Baxter County, Arkansas, 72635

Email: BCGI@CENTURYTEL.NET

- Background
- Nomenclature
- Introduction
- Coning Motion
- Assumptions and limitations of study
- Method of Studying the Coning Motion
- Aerodynamic Forces Acting on the Bullet
- Forces Driving the Coning Motion
- Mathematics of the Coning Motion
- Wind Shift Effects
- The Crosswind Aerodynamic Jump
- Yaw of Repose and Spin Drift
- The Coning Angle
- Energy Considerations
- Conclusions
- Summary

## Background

A spin-stabilized rifle bullet moves through the atmosphere in ballistic flight in accordance with the equations of motion which govern the flight path of any unpowered, unguided projectile together with its spin-induced gyroscopic reactions and their resulting aerodynamic effects.

By numerically integrating these controlling equations, a 6 degree-of-freedom digital simulation can provide realistically accurate calculations of a rifle bullet's attitude and trajectory. In linear aeroballistic theory the aerodynamic forces of drag and lift and the aerodynamic overturning moment are linearly modeled to enable accurate calculation of ballistic flight.

The ballistic trajectory is primarily determined by the aerodynamic drag force and acceleration of gravity and secondarily by any net lift forces acting on the projectile. The gyroscopic motions of the bullet's spin-axis are driven by the aerodynamic overturning moment acting upon that spinning bullet.

The well-developed Tri-Cyclic Theory provides an analytical tool for computing the epicyclic motions of the bullet's spin-axis attitude throughout its flight. This Coning Theory of Bullet Motions provides a similar analytical tool for computing the helical motion of the center of gravity (CG) of the spin-stabilized bullet about its mean trajectory during flight.

From analysis of simulated flight data, the motion of the CG of the bullet will be shown to be a damped, isotropic, right-circular conical, harmonic oscillation at the rate of gyroscopic precession with the nose of the spinning bullet angled inward toward its mean trajectory.

We term this the "coning motion" of the flying bullet. This helical harmonic motion of the CG of the bullet is driven primarily by its aerodynamic lift force and secondarily by its drag force. Coning parameters will be formulated in terms of traditional linear aeroballistic parameters and demonstrated to allow valid analyses of aerodynamic jump phenomena and the bullet's gradually increasing yaw-of-repose and spin-drift phenomena.

## Nomenclature
### Forces and Moments

$F_D = q*S*C_D =$ Aerodynamic drag force acting on the bullet (in pounds).

$F_L = q*S*Sin(\alpha)*C_{L\alpha} =$ Aerodynamic lift force acting on the bullet (in pounds).

$F = F_D + F_L =$ Total aerodynamic force acting on the bullet (in pounds).

$M = q*S*d*Sin(\alpha)*C_{M\alpha} =$ Aerodynamic overturning moment acting on the bullet (in pounds-feet).

$\alpha =$ Angle-of-attack of bullet spin-axis with respect to the apparent wind and also the magnitude of the half-cone-angle of bullet's coning motion (an angular value in radians).

$g =$ **32.174 feet per second per second** = Nominal acceleration of gravity.

$\rho =$ Density of the atmosphere through which the bullet is moving is often given in aeroballistics work in pounds (mass) per cubic foot. One pound mass (**lbm**) is the mass which produces one pound of force (**lbf**) in a gravitational field of **g** feet per second squared and has units of pounds-seconds squared per foot. The sea-level density of the standard ICAO atmosphere used here is **0.0764742 lbm/cubic foot** or **0.00237689 slugs/cubic foot**.

$q = \rho V^2/2 =$ Dynamic pressure (in pounds force per square foot) acting on the flying bullet.

$V =$ Bullet's mean airspeed velocity vector (in feet per second) in earth-fixed coordinates tangent to the mean trajectory. [$V$ is directed essentially horizontally in flat firing, and its magnitude is also essentially the supersonic ground-speed of the bullet.]

$S = \pi d^2/4 =$ Reference cross-sectional area of the bullet (in square feet) as viewed from ahead. the reference area $S$ is equal to **0.00051740 square feet** for a 30-caliber bullet.

$d =$ Barrel groove diameter = Caliber of the fired bullet (in feet).

$C_D =$ Coefficient of total aerodynamic drag as a dimensionless function of Mach-speed of the bullet and as an *even function* in $Sin(\alpha)$.

$C_{L\alpha} =$ Coefficient of lift as a dimensionless function of Mach-speed used in determining the lift force on the bullet as an *odd function* in $Sin(\alpha)$ when $\alpha$ is a signed angle-of-attack.

$C_{M\alpha} =$ Coefficient for determining the overturning moment acting on the bullet as a dimensionless function of Mach-speed of the bullet and an *odd function* in $Sin(\alpha)$ when $\alpha$ is a signed angle-of-attack. $C_{M\alpha}$ is the derivative of $M$ with respect to $\alpha$.

## Coning Motion Symbols

**r = D*Sin(α) =** Orbital radius (in feet) of the CG of the coning bullet revolving about a mean center of gravity moving along the mean trajectory with the bullet.

**$F_R$** = Centripetal Hookean restoring force (in pounds) needed to maintain a circular harmonic orbit at a given orbital radius.

**$F_C$** = Coning force component of the total aerodynamic force **F** perpendicular to the coning distance vector **D** (in pounds).

**$k_R$, $k_C$** = Slowly varying "force constant" values of restoring force per unit of radial displacement away from a neutral point at the center of the coning motion (in pounds per foot of displacement).

**m** = Mass of the bullet in slugs [**168/(7000*g)=0.000746 slugs** for our example **168-grain** bullet]. Ballisticians often use units of weight (pounds or grains) in non-metric aeroballistics work when giving the mass of a projectile. We will use mass in slugs here.

**v = r*$\omega_2$** = Circular orbital velocity of the CG of the coning bullet about its mean center (in feet per second).

**$T_2 = 2\pi/\omega_2$** = Inertial period (in seconds) of a coning cycle at the slow-mode gyroscopic precession rate $\omega_2$.

**$\omega_2 = 2\pi*f_2 = 2\pi/T_2$** = Circular frequency of gyroscopic precession, slow-mode (in radians per second) and also the circular coning frequency of the bullet.

**$\omega_1 = 2\pi*f_1$** = Circular inertial frequency of gyroscopic nutation or fast-mode epicyclic motion (in radians per second).

**$f_2$** = Inertial frequency of slow-mode gyroscopic precession (**65 hertz** initially in this example).

**$f_1$** = Inertial frequency of gyroscopic nutation, fast-mode (**311 hertz** initially in this example).

**D = $R_{CG}$ - $R_{Apex}$** = Distance vector from cone apex to CG of the coning bullet (in feet).

**$R_{CG}$, $R_{Apex}$** = Position vectors for CG of bullet and apex of cone, respectively, both in either earth-fixed or moving coordinates, as needed (in feet).

**$\Gamma_C$** = Rotating aerodynamic torque vector acting about the apex of the cone and driving the torsional coning motion of the CG (in pounds-feet). $\Gamma_C$ rotates with the coning bullet at $\omega_2$.

**β** = Angle whose trigonometric tangent is the instantaneous "lift-to-drag" ratio, $F_L/F_D$ (in radians).

**Δ** = Delta, the "small finite change in symbol following" operator. [**ΔX** is a vector if the symbol **X** represents a vector quantity.]

$I_C = m*D^2$ = *Coning moment of inertia* of coning bullet (considered as a *point mass* located at its CG) about the cone apex (in slug-feet squared).

$I_x = m*d^2*k_x^2$ = Moment of inertia of the bullet about its spin-axis (in slug-feet squared). [$I_x$ = 0.000247 lbm-inch$^2$ = 1.715x10$^{-6}$ lbm-ft$^2$ = 5.331x10$^{-8}$ slug-feet$^2$ for this bullet.]

$I_y = m*d^2*k_y^2$ = Moment of inertia about any transverse principal axis through the CG (in slug-feet squared). [$I_y$ = 0.001838 lbm-inch$^2$ = 1.2764x10$^{-5}$ lbm-ft$^2$ = 3.967z10$^{-7}$ slug-feet$^2$ for this bullet.]

$k_x, k_y$ = Radii of gyration about ($x, y$)-principal axes of the bullet given in calibers $d$.

$L = I_x*\omega$ = Vector angular momentum of spinning bullet about its spin-axis (in slug-feet squared per second).

$p$ = Vector spin-rate of the bullet about a principal axis aligned with its axis of symmetry (in hertz) = **2800 revolutions/second** (initially).

$\omega = 2\pi*p$ = Vector spin-rate of the bullet about its longitudinal principal axis of symmetry expressed as a circular frequency (in radians per second). [Note: $\omega$, $\omega_1$, and $\omega_2$ are the three cyclic rates of the Tri-Cyclic Theory.]

$\alpha(t)$ = Complex coning angle (in radians) as a function of time $t$.

$\varphi(t)$ = Aircraft-type pitch attitude of bullet spin-axis (in radians measured upward from $+V$ direction), the real part of $\alpha(t)$.

$\theta(t)$ = Aircraft-type yaw attitude of bullet spin-axis (in radians measured rightward from $+V$ direction), the imaginary part of $\alpha(t)$.

$i$ = Imaginary ($+\theta$) axis-direction unit vector in the complex plane. [$i^2 = -1$].

$K_0 = [\varphi_0^2 + (\theta_0 + \gamma)^2]^{1/2}$ = Initial magnitude of complex cone angle $\alpha_0$ (in radians).

$\xi_0$ = ATAN2$\{-\varphi_0, -(\theta_0 + \gamma)\} +\pi$ = Initial roll orientation or phase angle of the complex angle-of-attack $\alpha_0$, measured positive clockwise from the real $+\varphi$ axis (in radians from zero to $2\pi$). ATAN2$\{a_1, a_2\}$ is the two-argument ($a_1, a_2$) arctangent function used in FORTRAN.

## Apparent Wind and Transient Aerodynamic Effects

**W** = Instantaneous true wind vector in earth-fixed coordinates [**W**<<**V**] (in feet per second).

**W$_A$** = **W** – **V** = Apparent wind vector in a coordinate system moving with the flying bullet (in feet per second).

**γ** = Angular offset of apparent wind direction **W$_A$** from –**V** direction of the bullet due to crosswind **W** (in radians). [The offset angle **γ** is negative, in the -θ(t) direction, for the constant left-to-right crosswind used in this example.] The initial magnitude sum of the epicyclic sum of **γ** and **δ** is the initial magnitude of the slow-mode epicyclic arm.

**δ** = Angular magnitude of fast-mode nutating motion caused by sudden appearance of crosswind **W** (in radians). The absolute magnitude of **δ** is the magnitude of the fast-mode epicyclic arm.

**R** = Stability Ratio of epicyclic (inertial gyroscopic) rates (dimensionless). [$R = \omega_1/\omega_2 = f_1/f_2 = \varphi'_1/\varphi'_2 = 4.79 >> 1$, initially, in this example.] $R = 2*\{Sg + SQRT[Sg*(Sg – 1)]\} – 1$

**Sg** = $(R + 1)^2/(4*R)$ = Gyroscopic Stability of the bullet (dimensionless) = **1.75** initially, in this example. **R** is a more useful indicator of gyroscopic stability than is **Sg** itself.

**T$_N$** = Period of nutation with respect to the moving slow-mode arm = $2\pi/[\omega_1 – \omega_2] = 2\pi/[(R – 1)*\omega_2]$ = **4.06 milliseconds**, initially in this example.

**ξ$_1$, ξ$_2$** = Roll orientation arguments as functions of time (t) of fast-mode and slow-mode arms, respectively, of the epicyclic motion of the bullet's spin-axis about the apparent wind direction, each measured clockwise (in radians) from the **+φ** (pitch axis) direction.

**f$_N$** = $f_1 - f_2 = 1/T_N$ = **246 hertz** (initially) = Reduced (relative) nutation rate with respect to the moving slow-mode arm (in hertz).

**J** = Cross-track impulse due to transient aerodynamic coning force (in pound-seconds).

**A$_J$** = Aerodynamic jump angle (in radians), a permanent angular deflection of the flight path.

**ΔV$_C$** = Small cross-track "kick" velocity (in feet per second) due to impulse J.

**Φ** = Flight path angle (in radians) measured positive upward from local horizontal plane to a tangent to the mean trajectory projected into a vertical plane.

**β$_R$** = Yaw-of-repose angle of the bullet's coning-axis (in radians) measured positive rightward from launch azimuth to the axis direction of the coning motion. **β$_R$** is constrained to lie in the horizontal plane perpendicular to the local gravity gradient.

**β$_T$** = The Tangent Angle between the x-axis and a tangent to the mean trajectory projected into a horizontal plane.

**n** = Twist rate of rifled barrel in calibers per turn. [**n = 38.96** in this example.]

## Introduction

This paper puts forward a comprehensive *Coning Theory* explaining the motions of spin-stabilized rifle bullets in flight and provides insight into the nature of these motions. If not all new to the science of ballistics, at least these ideas are original in aggregate.

An early popular version of this theory was published in *Precision Shooting Magazine* [1] in 2011. Most basic is the concept that the center of gravity (CG) of the coning bullet always spirals around its mean trajectory at a larger radius than does the nose of that bullet.

We will show that the spinning bullet always "cones" around at its slow-mode, *gyroscopic precession* rate $\omega_2$ with its nose angled *inward* toward its *mean trajectory* as diagrammed in **Fig. 1**. The CG of the bullet revolves in the same rotational sense as the rifling twist around a mean center of gravity moving directly along its mean trajectory.

The projected direction of the bullet's spin-axis seen in the ballistician's customary wind axes orthogonal pitch-versus-yaw plot revolves in circular or spiral motion (in absence of epicyclic *gyroscopic nutation*) about the direction from which the apparent wind is approaching the flying bullet as shown at the right side of the diagram in **Fig. 1**. Thus, the axis of the bullet's coning motion is likewise continually defined by the apparent wind direction, except for two very small horizontal and vertical "tracking error angles."

Currently accepted aeroballistic theory seems to hold, instead, that the bullet should cone around with its nose pointing *outward*, and that the CG of the bullet itself should move directly along the trajectory.

For example, Harold R. Vaughn [2] has defined "coning motion" as:

"*The motion a bullet makes with its nose traveling in a circle while the CG remains fixed on the flight path.*"

This fallacy is cited only to illustrate the need for this new Coning Theory. While other working ballisticians might have disagreed with Mr. Vaughn, the situation has never been fully clarified.

Ballisticians define the "trajectory" or "flight path" of the CG of the bullet to include the "epicyclic swerve" of the bullet's coning motion. We instead define a *mean trajectory* which does not include these modulations. The *mean CG* of the coning bullet moves smoothly along this *mean trajectory* with its *mean velocity vector* always tangent to that *mean trajectory*.

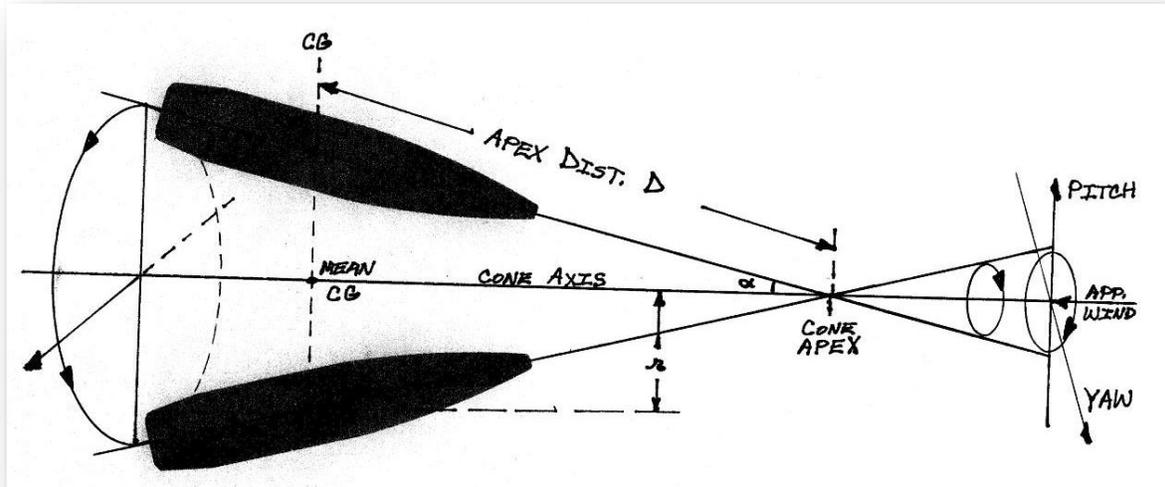

**Fig. 1**  Extreme Positions and Attitudes of Coning Bullet.

## Coning Motion

The *coning motion* of a spin-stabilized rifle bullet in free flight is a direct result of the slow-mode *gyroscopic precession* motion of its spin-axis which is driven by an aerodynamically produced overturning or pitching moment. As such, the pointing direction of the coning bullet's spin-axis follows a *circular, spiral, or epicyclic path (and never an elliptical path)* in the wind axes coordinate system plot shown on the right-hand side in **Fig. 1**. The gyroscopic precession of the bullet's spin-axis is *isotropic* in its rotational motion.

Likewise, the orbit of the CG about its mean trajectory is always a *right-circular helix* or *spiral* and is never *elliptical*. A non-zero slow-mode damping factor merely modifies the orbital radius and coning angle in a secular exponential manner with ongoing flight distance or time. If a higher-rate "fast mode" *gyroscopic nutation* is superimposed on this precession, the path of the bullet's spin-axis in the wind-axes plot becomes an epicyclic curve, and the spin-axis of the coning bullet simply wobbles around at this faster nutation rate without producing significant additional motion of its CG. One could reckon that a similar coning motion occurs at this faster nutation rate, but any additional motion of the CG is normally too small and too quickly damped to warrant any practical interest.

Current Tri-Cyclic ballistics theory explains these gyroscopic torque reactions, precession and nutation, perfectly well. According to this new Coning Theory, the motion of the CG of the free-flying bullet is a *right-circular conical isotropic harmonic oscillation* at the gyroscopic precession rate $\omega_2$ as shown on the left-hand side of **Fig. 1** in a coordinate system moving with the bullet along its mean trajectory.

This orbiting motion of the CG at the bullet's gyroscopic precession rate is primarily driven by the aerodynamically produced *lift force* and secondarily by the aerodynamic *drag force* as diagrammed in **Fig. 2**. The precession of the spin axis and this orbital coning motion of the CG are always perfectly synchronized with each other as will be developed in detail.

We theorize here that both the lift force $F_L$ and drag force $F_D$ contribute to the coning force $F_C$ driving the bullet's coning motion—as opposed to having only the lift force $F_L$ participating in driving this motion per the currently accepted analytical formulation of Robert L. McCoy [3] and possibly others at the US Army's former Ballistics Research Laboratory (BRL) at Aberdeen, Maryland. Perhaps they had been about to discover the true coning motion of spin-stabilized projectiles when Bob McCoy passed away unexpectedly.

This minor correction of incorporating drag in the coning motion is necessary for self-consistency of the Coning Theory and produces better agreement with 6-DoF simulations and, thus, with the empirical data. We frequently rely upon vector differential geometry in this study of the physics behind the bullet's coning motions in addition to the elegant vector calculus favored at BRL. These tools are sufficient to the task of illuminating more clearly the details of the coning motions.

Another important new concept is that the *axis of the bullet's coning motion*, and *not* the *spin-axis of the bullet itself*, as seems commonly to be believed, always points almost directly into the apparent wind approaching the bullet.

In fact, the direction of the apparent wind encountered by the bullet in flight *continuously defines the axis of the coning motion*. The very small vector angular momentum of the precession-rate coning motion itself can change its direction even more readily than can the ficklest of real winds. The moving spin-axis of the bullet quickly incorporates each change in the cone-axis direction into the coning motion itself—within less than one half of a coning cycle.

We should point out one important exception to this wind tracking ability of this coning motion, even though its occurrence is well outside the scope of this study: a spin-stabilized artillery projectile, nearing the apogee of a high-angle trajectory, might be unable to arc over quickly enough to continue its nose-forward coning motion during the descending limb of its flight. We say such a projectile has "failed to trail" as it falls to earth coning sideways, or even backwards, badly missing its intended target and perhaps failing to detonate [4].

Whenever the axis of the coning motion has to move in order to remain aligned with a new apparent wind direction, the only mechanism available for adjusting the coning motion is for the *cone apex* angle α to *increase in magnitude* to accomplish this re-alignment. As a harmonic motion, the coning rate of the bullet $\omega_2$ is independent of its coning amplitude α. The coning rate $\omega_2$ is determined by the $\omega_2$ gyroscopic precession rate of the bullet's spin-axis. The cone apex distance **D** remains fixed during any sudden wind shift.

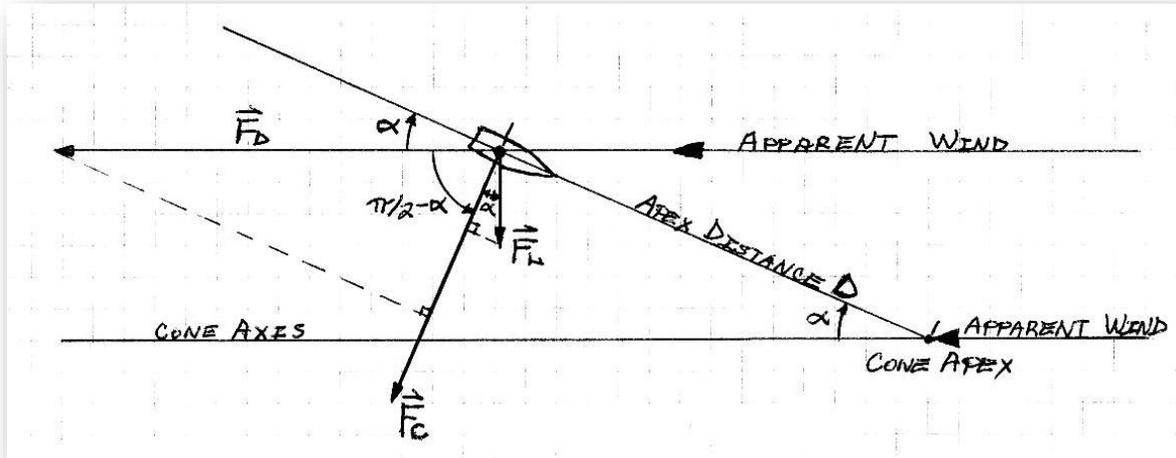

**Fig. 2  Powering the Coning Motion of the Bullet's CG**

The coning angle α of the bullet is also the magnitude of its aerodynamic angle-of-attack α relative to the approaching air-stream. The magnitude of the aerodynamic coning force **Fc** acting on the bullet increases *almost linearly* with any increase in angle-of-attack α and provides exactly the increased centripetal coning force **Fc** needed for the CG of the bullet to orbit around its mean trajectory at a new larger radius **r = D*Sin(α)**.

The increased yaw-drag force attributable the increased coning angle-of-attack needed to realign the coning axis into the new wind direction causes an additional loss in the kinetic energy of forward motion during the following half coning cycle. This loss in kinetic energy of the moving bullet matches the increase in the potential energy at the larger orbital radius of the CG of the coning bullet. Only an *increase* in the coning angle α can provide the needed energy to shift the coning axis direction.

In this way, the coning bullet is quickly able to align its cone axis, orienting it into any new apparent wind **Wa** by *increasing* its cone angle α and its corresponding coning radius **r**, even though the nose of the coning bullet itself is torqued away from the new wind direction by that approaching wind.

Only later in the flight, after the precession-rate coning motion due to that wind change has *almost damped out*, as it tends to do eventually for *dynamically stable* rifle bullets, will the spin-axis of the bullet be seen to have oriented itself into alignment with the new apparent wind direction.

The spinning bullet cannot just magically "turn its nose into the wind," and its overturning moment acts backward from one which could accomplish this alignment feat. The spinning bullet carries significant angular momentum. The coning motion of the spinning bullet is the magic which allows this quick average re-alignment of the nose of the bullet directly into the new apparent wind-stream $W_A$.

A one-time-per-firing transient coning motion, commencing when the spinning rifle bullet first encounters a new *purely horizontal* crosswind, explains the small and fairly recently documented *vertical-direction* crosswind aerodynamic jump in the flight path angle which can be reliably observed in precision shooting. There is also a horizontal crosswind aerodynamic jump which occurs in the same time interval, but this horizontal-direction trajectory deflection is rapidly cancelled by subsequent coning motion and produces no long-range impact error.

The non-cancelling vertical-direction trajectory deflection has been analytically formulated in calculus-based aeroballistic terms by Robert L. McCoy at BRL. Independent analytical calculations of this angular vertical deflection of the trajectory are presented here based on Coning Theory.

These numerical results agree fairly well with McCoy's values for our example bullet after the BRL formulation is adjusted to incorporate the small contribution of the bullet's drag force toward driving its coning motion.

Vertical deflections calculated from Coning Theory agree very well with the CG location bullet drop data from 6-DoF simulations of flat-firing in horizontal crosswinds. No Magnus effect of any type is involved in either formulation. A similar transient coning effect produces a similar type of one-time angular aerodynamic jump deflection of the trajectory whenever the bullet enters the windstream with a non-zero aeroballistic yaw (or yaw rate).

An incremental deviation from this nominal coning motion which recurs *twice* per coning cycle as the bullet's trajectory arcs ever downward due to gravity explains the slow increase in the yaw-of-repose of the bullet and, thence, the slow increase in the long-known spin-drift of the bullet in the same horizontal direction as the sense of the rifling twist.

This horizontal spin-drift of the CG of the bullet at long ranges is the accumulated net effect of the aerodynamic lift force acting rightward on the bullet due to a net left-side angle-of-attack of the small, but steadily increasing, rightward yaw-of-repose.

The center of the circular orbit of the bullet's CG always moves forward directly along (and defines) the mean trajectory. The apex of the coning motion of the CG always remains pointing almost directly from this orbital center location on the mean trajectory toward the "eye" of the approaching apparent wind $W_A$, and, thus, the coning apex does *not* actually lie directly upon the mean trajectory as it precedes any bullet which is flying through a crosswind or falling due to gravity. Even for flat firing in the complete absence of crosswinds, the coning axis for a right-hand spinning bullet points rightward of the tangent to the trajectory by a very small horizontal tracking error yaw angle $\varepsilon_H$ and above the tangent by a similarly small pitch-direction vertical tracking error angle $\varepsilon_V$.

On a macro-scale, the forward motion of the coning bullet is retarded by an average drag force aligned with the coning axis. In linear aeroballistics, this drag force can be calculated from the zero-yaw coefficient of drag **$CD_0$** augmented by a yaw-drag coefficient **$CD_\alpha$** the sine-squared of the coning angle $\alpha$.

The agreement of these detailed bullet motions with the existing *aeroballistic equations of motion* for bullet flight is demonstrated by the use of data outputs from existing six-degree-of-freedom (6-DoF) flight simulations, which numerically integrate these equations of motion, to show how this new Coning Theory explains the motions of rifle bullets in flight. With proper initialization, these 6-DoF simulator outputs can agree very well with Doppler radar and instrumented range measurements of the flights of real bullets.

The *Coning Theory of Bullet Motions* does not rely upon any of the minor aeroballistic forces or moments (spin-damping, pitch-damping, Magnus force or moment, etc.) in analytical explanation of these observed motions of spin-stabilized rifle bullets. No discussion of these non-relevant forces and moments is included here.

In this formulation of bullet coning motions, we need consider only the primary aerodynamic *forces* of *drag* and *lift* and the primary aerodynamic *overturning moment* acting on the bullet, all three of which combine at any given time during the bullet's flight into a single, instantaneous *total aerodynamic force*, a line-vector acting at one particular point on the surface of the bullet and with its line-of-action passing through the instantaneous center of pressure (CP) on the axis of symmetry, an inertial principal axis of the spinning bullet.

Coning motion *always occurs* in flat firing, even if a perfectly made bullet could be perfectly launched into a completely wind-free atmosphere.

This new Coning Theory generally coincides with and extends modern conventional analytic aeroballistic theory for spin-stabilized projectiles [4, 5]. This new Coning Theory should be in complete agreement with the precepts of classical mechanics [6], but the author must assume sole responsibility for any errors in its development.

## Assumptions and limitations of study

The modern Spitzer-style rifle bullet is a pointed, rigid, rotationally symmetric, statically unstable projectile of about 2.5 to 5.5 calibers in length. Spin-stabilization is applied to the bullet at launch to prevent its tumbling in flight. Throughout this study, the rotational sense of the bullet's spin is right handed, or clockwise as seen from behind the flying bullet. For purposes of this discussion these rifle bullets are assumed to have been perfectly manufactured in static and dynamic balance, profile shape, and axial symmetry. A principal inertial spin-axis of the fired bullet is assumed to coincide exactly with the mechanical axis of symmetry of its outside profile.

The rifle firing conditions being considered here are flat-firing with muzzle elevation less than 0.100 radians (5.7 degrees) from a firing position a few feet above a horizontal sea-level planar facet of the earth at 40 degrees, North latitude. The target is at the same elevation as the firing point. The firing direction is South-to-North, and a uniform (laminar flow) constant crosswind is moving at 10 miles per hour from West-to-East across the entire firing range. An ICAO-standard dry sea-level atmosphere at 59 degrees Fahrenheit and 29.92 inches of mercury (inHg) absolute barometric pressure is assumed. The speed of sound (Mach 1.0) is **1116.45 feet per second** and the density of this atmosphere is **0.0764742 pounds (mass) per cubic foot**.

Our subject rifle bullet, the obsolete 168-grain 30-caliber Sierra International bullet, is fired nearly horizontally at a target **900 yards** away. The rifle's sights are zeroed on the target center, and the line-of-sights is horizontal at firing. The maximum ordinate of the bullet's trajectory is only a few feet above the sight line. The barrel of our assumed target rifle is chambered in **308 Winchester** (or 7.62x51 mm NATO) and is rifled at a right-hand twist rate of **12 inches** (or **38.96 calibers**) per turn. The muzzle velocity used is **2800 feet per second**. Our subject bullet is engraved perfectly by the rifling during firing and emerges from the muzzle-blast cloud with **zero** initial aeroballistic yaw and yaw-rate.

For a rotationally symmetric projectile, its aeroballistic yaw can be thought of as the Root-Sum-Square (RSS) of its small, non-Eulerian, orthogonal "aircraft-type" *pitch* (up or down) and *yaw* (side-slip) attitude angles. This generalized aeroballistic yaw is also the magnitude of the aerodynamic angle-of-attack for rotationally symmetric projectiles as used herein.

The very real flight-disturbing effects of bullet static and dynamic imbalance, in-bore yawing, muzzle-blast and muzzle brake disturbances, or the motions of the muzzle of the recoiling rifle barrel (for examples) are not discussed here.

Our subject rifle bullets are further assumed to be statically unstable, but spin-stabilized for gyroscopic stability (initial **Sg > 1.5**), and also assumed to be dynamically stable throughout all or most of their almost horizontal supersonic flights in the flat-firing case being studied here. We are considering only the supersonic portion of flat-firing rifle bullet trajectories in which the forward speed of the bullet is monotonically decreasing. In high-angle firing, the projectile sometimes accelerates due to gravity during the downward leg of its trajectory.

By saying that a rifle bullet is "statically unstable," we mean that the aerodynamic center of pressure (CP) for the modern rifle bullet flying normally at small yaw angles is *ahead* of its center of gravity (CG). Both centers are located on the axis of symmetry of this ideal rotationally symmetric bullet. Our selected example bullet, the well-studied 30-caliber 168-grain Sierra *MatchKing* (formerly *International*), is launched with an initial gyroscopic stability ($S_g$) of **1.75.**

This example rifle bullet is just slightly unstable dynamically. That is, the angular amplitude of its coning motion **α(t)** slowly increases exponentially with flight distance **s** or flight time **t** rather than damping down exponentially as with most rifle bullets. This unusual flight behavior makes this selected example bullet particularly useful for the study of coning motion.

The long axis of the rifle bullet is also a *principal axis of inertia*, producing an *extremum* (either a *minimum* or a *maximum*) in the second moment of the mass distribution of the bullet about that axis. In the case of a rifle bullet, its spin-axis is the axis having a *minimum* moment of inertia; i.e., the axial direction producing the smallest possible second moment of the mass distribution of that bullet.

For our example 30-caliber rifle bullet, the moment of inertia about its spin axis has a *minimum* possible value **1/7.44 times** than that about any transverse principal axis through the CG. A longer 7 mm bullet, for example, of this same weight and type would have an even larger ratio of its second moments of inertia.

For a conventional wheel-type gyroscope in which the mass distribution can be resolved into ring-shaped elements in a thin disc, the spin-axis has a *maximum* possible moment of inertia of just ***twice*** that about any transverse principal axis, for an **Iy/Ix** ratio of **1:2**. The spinning rifle bullet is simply a gyroscope of a different flavor. By instead spinning the bullet about a principal axis having a minimum second moment of inertia, we can readily produce solid monolithic rifle bullets having **Iy/Ix** ratios up to **14:1**.

Of course, it is more convenient to simulate and study these perfectly launched, ideal rifle bullets rather than dealing mathematically with the definition and physical effects of the many flaws to which real bullets fired from actual rifles are prone. However, we make these simplifying assumptions here because: 1) they make our studies easier to conduct and comprehend; and, 2) the study of this idealized case might actually be the stronger form of analysis in this instance. This new Coning Theory does *not* rely upon the presence of small imperfections in the projectile as do, for examples, the *Tri-Cyclic Theory for Missiles Having Slight Configurational Asymmetries* of J. D. Nicolaides of BRL **[7]** and the x- and y-spirals theorized much earlier by Dr. Franklin W. Mann **[8]**.

## Method of Studying the Coning Motion

To characterize this coning motion, one can examine streams of numerical flight simulation data values calculated on small uniform time intervals in a 6-DoF digital simulation of the flight of a perfectly launched, ideal example of our selected 30-caliber rifle bullet, the 168-grain Sierra *MatchKing*. We have the necessary aeroballistic coefficient data to perform these simulations for this bullet, at least in the guise of its substantially similar ancestor, the old 168-grain Sierra *International* bullet [9].

These aeroballistic coefficients are tabulated as functions of the Mach-speed of the bullet, the ratio of the speed of the bullet through the air to the velocity of propagation of acoustic pressure waves traveling through the specified atmosphere. We use a rather dense, dry, sea-level ICAO standard atmosphere at 15 degrees Celsius (59 degrees Fahrenheit) throughout these studies to assure reasonably large aerodynamic effects will be available for study.

Computed *bullet drift* and *bullet drop* data streams reported on **0.2 millisecond** time centers were analyzed to study the *horizontal* and *vertical* components, respectively, of the bullet's CG location as seen in an *earth-fixed coordinate system*. In addition to the possible precession and nutation rate motions of the bullet's CG, the *horizontal drift* (in inches) of the bullet's CG includes the crosswind-drift, horizontal Coriolis effect, horizontal aerodynamic jump components (angular trajectory deflections), and spin-drift.

The *vertical drop* (also in inches) below the projected axis of the bore, while mostly due to the acceleration of gravity, also includes the effects of any vertical crosswinds, the vertical component of the total aerodynamic force acting on the bullet, the vertical Coriolis effect, as well as any vertical direction aerodynamic jump experienced by the bullet such as the initial crosswind aerodynamic jump.

Coning of the CG at the gyroscopic precession rate **$f_2$ hertz** and any measurable CG motion occurring at the nutation rate **$f_1$ hertz** are the only possible *periodic modulations* of these two orthogonal CG-location data streams.

Each of the other variations produces only a slowly varying *secular* (non-periodic) effect. All of these analytic effects are boiled together into these two streams of *non-analytically produced* uniform-time-series CG-location values as output from the 6-DoF simulator.

A time-symmetric (non-causal), unit power, low-frequency-passing, digital filtering technique is employed in the temporal domain to remove all periodic modulations from the two data streams, leaving only the extraneous low-rate secular variations noted above, without time-shifting or distorting the amplitude of the remaining data values.

The low-pass filter is designed to have a sharp cut-off at a frequency matching the gyroscopic precession rate $f_2$ of the spinning bullet in a particular selected early portion of its flight so that all precession and nutation-rate modulations are removed. This filtering technique is designed to recover the **mean trajectory** of the CG of the bullet during its simulated flight.

The low-pass-filtered mean trajectory data arrays are then *subtracted*, point-by-point, from the original arrays of CG location data calculated by the 6-DoF simulator. The two data arrays remaining after this low-pass filtering and differencing procedure contain only the unmodified, full amplitude, precession-rate $f_2$ (and higher-frequency) modulations of the path of the bullet's CG.

The period $T_2$ (initially **15.4 milliseconds**, here) of the coning motion of the spinning bullet (i.e., the reciprocal of $f_2$, its slow-mode precession rate in hertz) is used in the design of a time-symmetric (non-causal), unit-power digital filter to extract the $f_2$-rate (**65 hertz**) modulation from each of the two uniform-time-series data streams. An *equally weighted running mean of one period* ($T_2$) *in length* was selected for the type of *convolutional digital filter* to be applied to these time-domain data streams. This filter operator spans a time interval of:

$$2*N*\Delta t \approx T_2 \geq 15.4 \text{ milliseconds} \tag{1}$$

where **N** is a small positive integer (initially **N = 39**) and $\Delta t$ is the fixed data sample interval (**0.2 msec**). The integer half-length **N** of the filter increments occasionally as the period $T_2 = 1/f_2$ lengthens gradually over the flight. In adapting the filter half-width **N** to the slowing coning rate $f_2$, always round up the value **N** to assure that the filter removes all the modulation. The circular coning rate $\omega_2 = 2\pi*f_2$ slows continually throughout the bullet's flight because 1) the ratio $R = f_1/f_2$ *increases* with increasing gyroscopic stability (**Sg**), and 2) the **sum** ($f_1+f_2$) *decreases* proportionally with the bullet's spin-rate **p** as it slows exponentially with time (**t**). The net result is a gradual decrease in the coning rate $\omega_2$ for the usual rifle bullet in supersonic flight.

At each position of the moving convolutional filter, the low-pass-filtered average value is subtracted from the original data sample aligned with the center of the **(2*N+1)**-point running mean. The first and last **N** data points are not available as filtered values due to end-effects of the filter operator itself.

The remaining data stream contains only the modulating frequencies of $f_2$ hertz (and possibly higher). The details of this procedure are included here to assist others in replicating or varying this experiment using their own data sets.

For convenience of analysis, the aircraft-type orthogonal pitch and yaw attitude angles of the right-hand spinning bullet's spin-axis in the wind-axes plot, and as tabulated for each sample time in the data streams, are converted into *polar coordinates* centered on the *apparent wind direction* (**0, i*γ**). The extracted coning-modulated drift and drop data streams might also be converted into polar coordinates about a mean CG origin moving along the mean trajectory.

Examination of these four resulting tabulated numerical data streams yields the following observations:

*The residual modulation of these two drift and drop CG position data streams shows that the periodic rate of the circular coning motion of the bullet's CG around its recovered mean trajectory matches the gyroscopic precession-rate ($f_2$ hertz) of the bullet's spin-axis, but that the CG of the bullet rotates exactly 180 degrees out-of-phase with the precession of the bullet's spin-axis at $\omega_2 = 2\pi*f_2$ radians per second as seen in wind axes plots.*

This relative phasing of the synchronous coning motions of the bullet's CG and gyroscopic precession of its spin-axis directions can only be consistent with an *inwardly angled* orientation of the bullet's nose which, in turn, could occur only if the CG of the bullet were located behind a crossing point (i.e., a cone apex) moving along (but not necessarily directly upon) the mean trajectory ahead of the bullet, as shown earlier in **Fig. 1**.

In this simulated flight, the coning motion is initiated by the perfectly launched, non-coning, simulated bullet encountering a constant left-to-right **10 mile-per-hour** horizontal crosswind at launch immediately after penetrating the envelop of the muzzle-blast cloud.

A nutation-rate wobbling motion due to suddenly hitting that crosswind is superimposed upon the precession-rate (**65 hertz**) coning motion of the bullet's spin-axis direction. The presence of this nutation is actually required for Conservation of Angular Momentum across this (or any other) step-change in flight conditions. This high-rate (**311 hertz**) nutating motion damps to imperceptibility after 5 or 6 additional coning cycles for our example bullet and does not produce any detectable CG motion, even while undamped. We designed a similar digital filter to isolate any modulation of CG position at the **311-hertz** nutation rate, but none could be detected in the undamped portion of the data set for this example bullet. The secondary coning of the CG at the nutation rate is negligible for well stabilized practical rifle bullets.

Wind-axes plots of all precession-rate motions of the bullet's spin-axis pointing direction from 6-DOF simulator runs show *circular*, or at most slowly inward or outward *spiraling*, wind-centered coning motions completely lacking any hint of *ellipticity*. The calculated CG motions are similarly *circular* or slowly *spiraling* around the mean trajectory.

The centers of the circular or slowly spiraling pitch- and yaw-coordinate values in the observed wind-axes plots consistently indicate that the coning bullet's spin-axis always revolves about the *instantaneous apparent wind direction*, which further indicates that the axis of the bullet's coning motion always points *directly into the instantaneous apparent wind*, except for some minor dynamic lagging during changes in that apparent wind direction.

## Aerodynamic Forces Acting on the Bullet

For certain bullets [10], we have tables of linear aeroballistic coefficients as functions of the Mach-speed of the bullet in flight which allow us to calculate the total aerodynamic force **F** experienced by the bullet at any point in its flight in terms of the magnitudes of its rectangular components, the drag force **F$_D$** and lift force **F$_L$**, as functions of the airspeed **V** of the bullet, the density ρ of the atmosphere, and the bullet's angle-of-attack α:

$$\{F_D\} = q*S*CD \qquad (2)$$

$$\{F_L\} = q*S*Sin(α)*CLα \qquad (3)$$

Then, the total aerodynamic force **F** acting on the bullet through its Center-of-Pressure (CP) is defined by the rectangular vector-summing relationship:

$$F = F_D + F_L \qquad (4)$$

Variations in the density ρ of the local atmosphere are handled by incorporating that variable directly into the formulation for the dynamic pressure **q**. Variation in the elasticity, density, and viscosity of the local atmosphere with ambient temperature, pressure, and humidity, and hence variation in the "speed of sound," are handled by tabulating the linear aeroballistic coefficients, **CD** and **CLα**, as functions of the bullet's Mach-speed **M** instead of its airspeed **V**.

Our example Spitzer-style rifle bullet is typical in that its transonic instability limits our study to airspeeds above about Mach 1.2 where steady supersonic flight can be maintained. In the bullet speed range of interest here (from a muzzle velocity of Mach 2.5 down to Mach 1.2), the supersonic drag is not really quite proportional to the square of velocity, as incorporated into the (subsonic) formulation of the dynamic pressure **q**.

Instead, the bullet's supersonic drag is approximately proportional to the 3/2-power of its airspeed, and this difference explains most of the variance in the tabulated drag coefficient **CD** as a function of Mach-speed above Mach 1.2.

The drag force **F$_D$** always acts in a *downwind direction* along the direction of relative motion of the undisturbed local air-mass as seen from the moving bullet; i.e., in the direction of movement of the apparent wind **W$_A$** approaching the bullet. The *direction* of the drag force vector **F$_D$** is independent of the *orientation* of the bullet in flight. The *magnitude* of the bullet's drag coefficient **CD** is found from the sum of a tabulated primary zero-yaw drag coefficient **CD$_0$** function of Mach-speed and a tabular additive adjustment for each Mach-speed of yaw-drag coefficients **CDα** to be multiplied by δ$^2$ (the square of the sine of α, the angle-of-attack of the bullet). Thus, aerodynamic drag remains an *even function* of the angle-of-attack α.

The aerodynamic lift force **F$_L$** is defined to act *perpendicularly to the drag force* **F$_D$** and is thus constrained to act in a *plane perpendicular to the instantaneous apparent wind direction*. The aerodynamic lift force is an *odd function* of the angle-of-attack α in linear aeroballistics. In the case of simple coning motion, this lift force is directed radially inward toward the axis of the coning motion of the bullet. The instantaneous roll orientation of the lift force vector $F_L$ for some non-zero angle-of-attack $\alpha$ is completely determined by the instantaneous orientation of the plane containing the spin-axis of the bullet and the eye of the apparent wind, with that angle-of-attack $\alpha$ being the angle between those two directions. The "eye of the wind" is an ancient nautical expression for the exact *direction* from which the apparent wind is blowing at any instant as experienced aboard a moving vessel. Historically, ballisticians have sometimes imprecisely defined the direction of the lift force as acting perpendicularly to the trajectory of the projectile instead of perpendicularly to the direction of the instantaneous drag force. In linear aeroballistics, the coefficient of lift $CL\alpha$ is tabulated as a function of Mach-speed, but is itself virtually independent of the angle-of-attack $\alpha$ for our particular example rifle bullet, and needs no additional adjusting coefficient.

From study of the *statics of rigid bodies*, we can state the following three theorems about the complete system of aerodynamic forces acting on the free-flying bullet at any instant:

I. Disregarding any possible aerodynamic effect of the spinning of the bullet and treating the bullet as a non-rotating rigid body, we can sum the entire system of aerodynamic pressures and frictional (shear) forces acting over the whole surface of the bullet into a *single*, total aerodynamic force line-vector $F$ acting at the one point on the surface of the bullet which uniquely produces the instantaneous overturning moment $M$ which is also being experienced by the bullet. [No direct aerodynamic spin effects are apparent for our example bullet flying at small angles-of-attack $\alpha$ in this supersonic airspeed regime.]

II. Because 1) the shape of the bullet is effectively a closed-ended, axisymmetric solid of revolution, 2) the aerodynamic force $F$ is a line-vector which produces a torque on the bullet, and 3) the bullet is a rigid body; we can translate this total aerodynamic force line-vector $F$ from the surface of the bullet, near its nose, along the line-of-action of the line-vector $F$, which must intersect the bullet's spin-axis by symmetry, to a Center-of-Pressure CP lying on that spin-axis. [The force $F$ still produces the same overturning moment $M$ when it is applied at the axial CP location of the bullet.]

III. We can once again translate the force vector $F$ rearward from the CP along the spin-axis to the center-of-gravity CG of the bullet if we also separately consider the overturning moment $M$ as being produced by the resulting *force couple* ($F$, $-F$) acting on the whole bullet as a rigid body, but considered as a torque vector acting about the CG of the bullet. [The original force $F$ of the couple always remains acting through the CP of the bullet, while the added force $-F$, completing the couple, acts at the CG of the bullet exactly compensating the newly-added *translated* force $F$ now also acting at the CG.]

By this procedure, we de-couple the translational and overturning aerodynamic effects so that these can be analyzed separately. The location of the CP might well migrate along the spin-axis of our example bullet as the flight progresses and its airspeed and coning angle-of-attack $\alpha$ gradually change with ongoing time-of-flight. In addition, we should remember that

19 / 59   A Coning Theory of Bullet Motions – Boatright – rev. March/2018

changes over time in the total aerodynamic force **F**, together with its lift **F**$_L$ and drag **F**$_D$ component forces (which will be shown to drive the orbital coning motion of the CG) and its associated overturning moment **M** (which drives the Tri-Cyclic gyroscopic precession and nutation motions of the spin-axis), always remain *perfectly synchronized*. Of course, this synchronicity is because they each are just different manifestations of the common total aerodynamic force line-vector **F** at any instant in time during the flight. We will consider the overturning moment **M** in later sections.

## Forces Driving the Coning Motion

If the CG of the bullet is to orbit about its mean trajectory at its gyroscopic precession-rate $\omega_2$ as an *isotropic harmonic oscillation*, it must move as if the bullet were subject to some hypothetical, radially symmetric, centripetal restoring force **F**$_R$ which is *linearly proportional* to the radius **r** of this orbit and takes the form given by Hooke's Law as:

$$F_R = -k_R * r \quad (5)$$

Furthermore, from Newton's *Second Law of Motion*, for a body moving in a *circular orbit* at this *orbital rate* $\omega_2$ in such a force field, this unspecified centripetal restoring force **F**$_R$, whatever its source, must also equal:

F$_R$ = (Bullet Mass)*(Centripetal Acceleration for a Circular Orbit)

$$F_R = m*(-v^2/r) = -m*(\omega_2*r)^2/r = -m*\omega_2^2*r = -k_R*r \quad (6)$$

More precisely, the isotropic coning motion of the CG of the rifle bullet is a type of torsional harmonic oscillation in cone angle $\alpha$ about a cone apex which remains a relatively fixed distance **D** ahead of the CG of the bullet in flight. One could envision the CG of the bullet being affixed as a point mass at the tip of a massless fly-rod of length **D**. Earlier analysis **[3]** had the CG of the bullet moving in response to a simple radial restoring force, the aerodynamic lift force **F**$_L$ alone, incorrectly stated to be acting perpendicularly to the trajectory of the bullet, with *no* contribution from the drag force **F**$_D$ toward driving the coning motion.

We can define a vector **D**, giving the position of the CG of the bullet relative to the apex of the cone in any suitable coordinate system by the vector relationship:

$$D = R_{CG} - R_{Apex} \quad (7)$$

Then, as diagrammed in **Fig. 2** above, we can formulate an aerodynamic torque vector $\Gamma_C$, driving the torsional coning oscillation about the apex of the cone, as the vector cross-product:

$$\Gamma_C = D \times F = D \times F_D + D \times F_L$$

or, in *magnitudes*

$$\{\Gamma_C\} = D*F*\sin(\alpha+\beta) = D*F_C \tag{8}$$

where the magnitude of the *coning force* $F_C$ is the total size of the aerodynamic force component acting *perpendicularly* to **D** and thus directly available to drive the orbital coning motion in torsion. The angle between the vectors **D** and **F** is $\alpha+\beta$, where $\beta$ represents the small angle whose tangent is the lift-to-drag ratio $F_L/F_D$ of the bullet flying through the specified atmosphere at airspeed **V** and with angle-of-attack $\alpha$:

$$\beta = \tan^{-1}[F_L/F_D] = \tan^{-1}[(CL\alpha/CD)*\sin(\alpha)] \tag{9}$$

Thus, the lift-to-drag angle $\beta$ will normally exceed the angle-of-attack $\alpha$ by a significant ratio.

If the aerodynamic driving force $F_C$ is perpendicular to the direction of **D**, then from trigonometry the total force vector **F** projects in this perpendicular direction as:

$$\{F_C\} = F*\sin(\alpha+\beta) = [F*\sin(\beta)]*\cos(\alpha) + [F*\cos(\beta)]*\sin(\alpha)$$

$$\{F_C\} = F_L*\cos(\alpha) + F_D*\sin(\alpha) \tag{10}$$

For these small coning angles ($\alpha < 0.10$ radians = 5.7 degrees), we can approximate:

$$\cos(\alpha) \approx 1.00$$

Then, after this simplification and substitution of the aeroballistic expressions from **Eq. 2** and **Eq. 3** for the components of the force vector **F** into the expression for $F_C$ in **Eq. 10** above, we see an interesting and fundamental relationship defining the *magnitude* $\{F_C\}$ of the coning force $F_C$ available to drive the coning motion:

$$\{F_C\} = q*S*\sin(\alpha)*[CL\alpha+CD] \tag{11}$$

So, the drag force $F_D$ does indeed contribute somewhat (about 10 to 20 percent in most cases for small coning angles) toward driving the coning motion of the rifle bullet along with the more direct contribution from the smaller lift force $F_L$. In the absence of any fast-mode nutation, the bullet's coning force $F_C$ would be equivalent to the "normal force" acting upon a lifting body as sometimes used in aerodynamics.

Since $\sin(\alpha)$, the trigonometric sine of the half-cone-angle $\alpha$, can also be expressed geometrically as the ratio $r/D$, and adopting our negative sign convention for a centripetal force, we can put this expression into the form of a Hookean restoring force by those changes:

$$\{F_C\} = -q*S*\sin(\alpha)*[CL\alpha+CD] = -[q*S*(CL\alpha+CD)/D]*r = -k_C*r \tag{12}$$

Furthermore, if the aerodynamic coning force $F_C$ on the bullet is actually to provide the hypothetical centripetal force $F_R$ necessary to maintain this circular harmonic orbit, then at any given time these two force constants $k_R$ (from **Eq. 6**) and $k_C$ above, must be *equal* (at least for $\alpha < 5.7$ degrees), so that, solving for the cone apex distance **D**, in feet, we have another important and fundamental coning relationship:

$$k_R = m*\omega_2{}^2 = k_C = q*S*(CL\alpha+CD)/D$$

$$D = q*S*(CL\alpha+CD)/(m*\omega_2{}^2) \qquad (13)$$

and, for any given coning angle $\alpha$, the corresponding coning radius **r** can then be computed as:

$$r = D*Sin(\alpha) = q*S*Sin(\alpha)*(CL\alpha+CD)/(m*\omega_2{}^2) \qquad (14)$$

Thus, we have an expression for the cone apex distance **D** in **Eq. 13** as a function of several invariant or slowly varying aeroballistic parameters. The cone apex distance **D** is a fundamental parameter describing the bullet's coning motion. The coning bullet adjusts its apex position, and hence its apex distance **D**, as its aeroballistic parameters change slowly during its flight. The cone apex distance **D** starts out at about **1.25 inches** (or four calibers) just after launch for our example 30-caliber rifle bullet and gradually increases to about **3 inches** (or ten calibers) at **900 yards** downrange.

The coning radius **r** is approximately **0.10\*D** for $\alpha$ = **0.10 radians** (5.7 degrees), so **r** itself is not often much larger than about **0.3 inches** (or one caliber). The distance **D**, from the apex of the cone to the CG of the bullet, serves effectively as a lever arm, converting the forces driving the coning motion into a net torque $\Gamma_C$ about the cone apex driving the torsional harmonic oscillation of the bullet.

The cone angle $\alpha$ itself, with its corresponding coning radius **r**, is a *free variable* in this precession-rate oscillation and is thus available to increase (*only*) in size to accommodate almost instantaneously each change in flight conditions which might be encountered by the free-flying bullet. Except for gradual slow-mode damping for dynamically stable bullets, the size of the coning angle $\alpha$ (and its corresponding coning radius **r**) is constrained to be non-decreasing in response to any changes in flight conditions which might be encountered. Any step-change in flight conditions which might seem able to *decrease* the coning angle by some amount for a bullet at one point in its coning cycle will also *increase* the coning angle by *twice as much* for a bullet positioned opposite that location in the coning motion. The *increase* in coning angle $\alpha$ always controls the coning motion of the bullet.

As a simple way of envisioning this coning response to a step-change in flight conditions, imagine the CG of the bullet moving in a circle about the mean trajectory in a perpendicular plane moving with the flying bullet. If the orbital radius of the CG is **r**, and the center of the circle were displaced by $\Delta r$ in some radial direction, the maximum eccentric radius is **r + $\Delta r$**. If the circular orbit is to continue around at this new, larger radius (**r + $\Delta r$**), the opposite side (shortest radius = **r - $\Delta r$**) must increase by **2\*$\Delta r$**. Now, think of these changes in terms of coning angles $\alpha$ instead of radii with **r = D\*Sin($\alpha$)**. Of course, the mean CG location always follows along the mean trajectory.

Another way to understand why the coning angle-of-attack of the bullet can only *increase* in response to any step-change in its approaching apparent wind comes from considering the small orbital energy of the coning motion itself and the much larger kinetic energy of the bullet due to its forward motion. For a circular orbit such as our isotropic harmonic oscillation of the bullet's CG here, the orbital potential energy and the orbital kinetic energy are always equal. The increased orbital potential energy for a larger orbital radius must be matched by an equal increase in the orbital kinetic energy to achieve that larger radius.

The coning bullet must do work to shift the axis-direction of its coning motion. That work (energy) can only be extracted from the huge kinetic energy of the bullet's forward motion by momentarily increasing the aerodynamic drag force retarding that forward motion. In linear aeroballistics, the instantaneous drag force is always determined by environmental and Mach-speed parameters except for the bullet's coning angle-of-attack $\alpha$. The yaw-drag force component, $q*S*Sin^2(\alpha)*CD_\alpha$, can increase only if $\alpha$ is *increased*.

Because the orbital radius $r = D*Sin(\alpha)$, the required increase in orbital potential energy is some small fraction **e** of the extra kinetic energy extracted by increasing the bullet's yaw drag with $Sin^2(\alpha)$ after a trajectory distance traversed by the bullet during *one half the current period of a coning cycle*, which is also the response time for the coning motion's tracking of shifts in approaching apparent wind.

## Mathematics of the Coning Motion

The oscillation in cone angle α about the cone apex can be described mathematically for a spin-stabilized bullet flying through a crosswind in terms of a *complex cone angle* α(t), having *real* **pitch** φ(t) and *imaginary* **yaw** θ(t) orthogonal aircraft-type attitude-angle components as functions of ongoing time **t**:

α(t) = φ(t) + i*[θ(t) - γ]                                                           (15)

Here, **γ** as a signed variable is a negative, leftward, angular yaw offset of the incoming direction of the apparent wind $W_A$ from the origin of the wind axes while the bullet is experiencing a left-to-right horizontal crosswind. No similar pitch offset is shown because of the definition of the origin direction of the standard wind axes plots being always in the **+V** direction, and because we are not studying the effects of vertical crosswinds here.

To the extent that the pitching over of the coning bullet follows the change in vertical flight path angle **Φ** (the tangent to the mean trajectory in a vertical plane) during the flight, this change in pitch attitude is invisible in wind-axes plots. The same tracking ability should also cause the yaw-of-repose to be similarly invisible in wind axes plots. Only the vertical and horizontal direction *tracking error angles* ($ε_V$ and $ε_H$) should show up in very small scale (auto-scaled) wind axes plots for no-wind zero yaw, zero yaw-rate 6-DoF simulations. Interestingly, the downward curving of the trajectory flight path angle due to gravity *leads* the coning-axis attitude response slightly, but the rightward curving yaw-of-repose attitude angle $β_R$ *leads* the rightward curving **+V** direction of the mean trajectory. These are two different stimulus/response relationships with the aerodynamic responses always slightly lagging the driving stimuli. The time-constant for each stimulus/response relationship is one half the period of the coning motion.

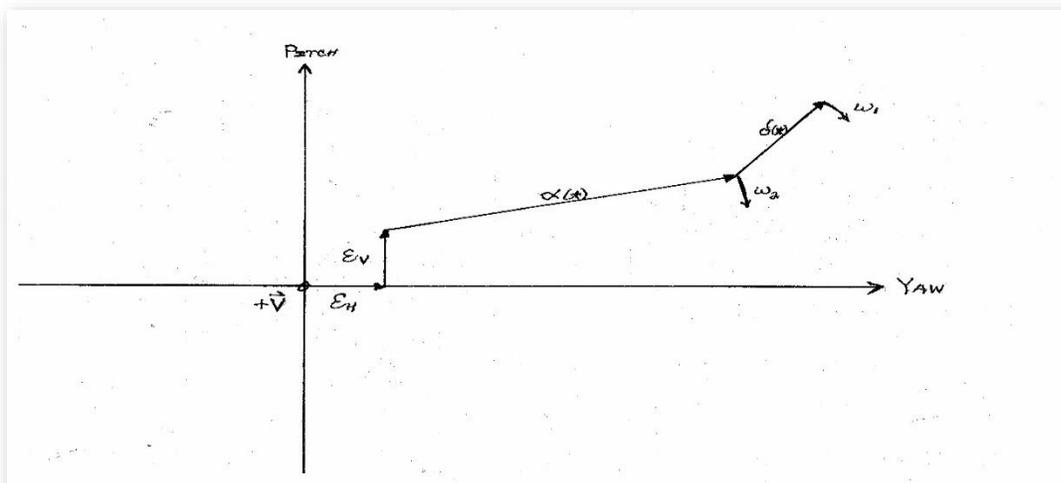

**Fig. 3**   Wind Axes Plot with No Atmospheric Wind

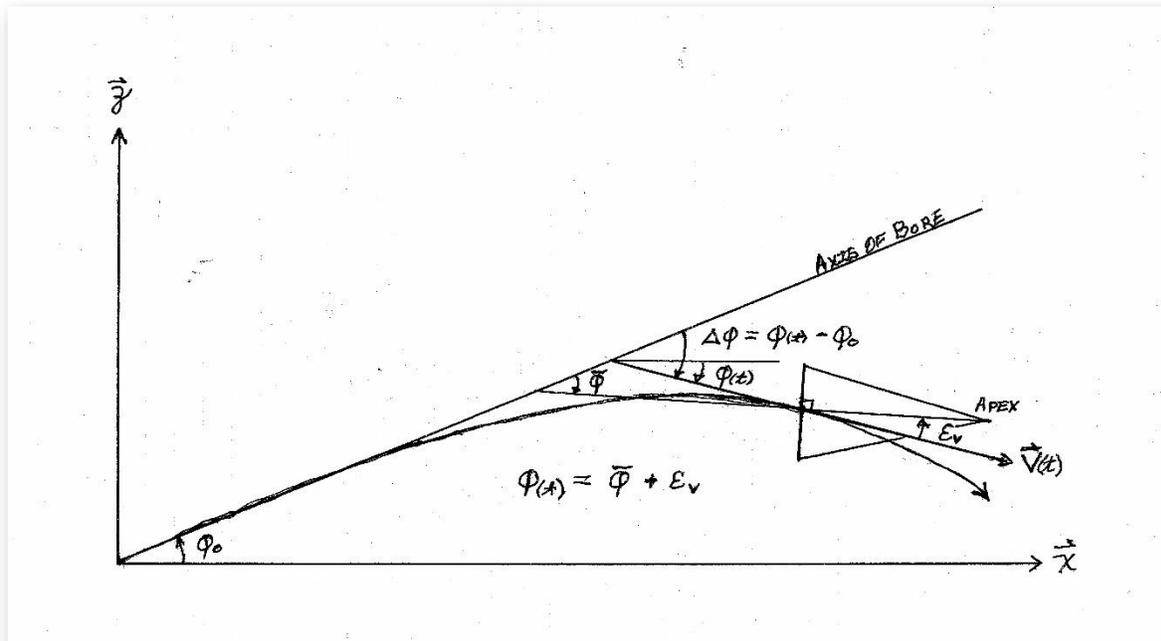

Fig. 4  Coning Motion in Vertical Plane View

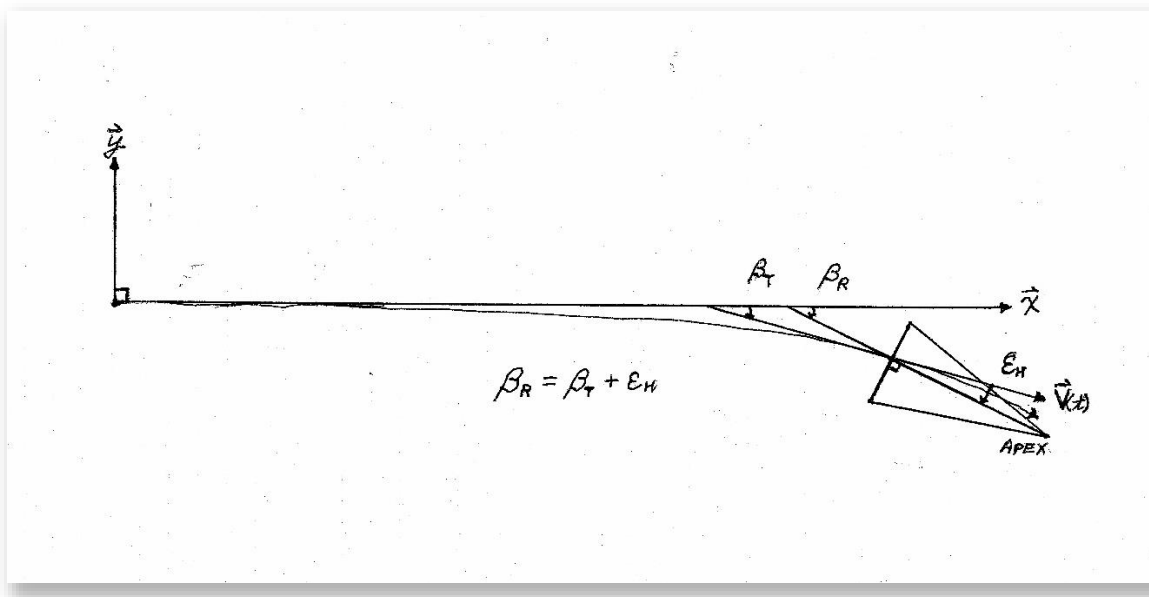

Fig. 5  Coning Motion in Horizontal Plane View showing RH Spin Drift (No Crosswind)

 A Coning Theory of Bullet Motions – Boatright – rev. March/2018

Making use of a torsional version of Newton's *Second Law of Motion* and **Eq. 8** above, we can write the magnitude relationship:

$$\Gamma_C = I_C * d^2\alpha/dt^2 = D * F_C \tag{16}$$

Substituting our previous expressions for the magnitudes of $I_C$ (formulated here as $m*D^2$, considering the bullet as a point mass), **D** (from **Eq. 13**), and $F_C$ (from **Eq. 12**), and invoking the same small-angle approximation which we use in reducing a similarly slightly non-linear expression to the form of Hooke's law when treating the small-amplitude motion of a simple pendulum as simple harmonic motion, we find that:

$$d^2\alpha/dt^2 = -\omega_2^2 * \alpha(t) \tag{17}$$

A typical solution in real and imaginary parts for this second-order ordinary differential equation in complex **α(t)** can be expressed as an orthogonal pair of the well-known relationships for circular harmonic oscillation at the precession rate $\omega_2$:

$$\varphi(t) = K_0 * Cos(\omega_2 * t + \xi_0)$$

$$\theta(t) = K_0 * Sin(\omega_2 * t + \xi_0) \tag{18}$$

With **(K₀, ξ₀)** and **(K₀, ξ₀-π/2)** taken as arbitrary constants of the four integrations.

These two parametric equations, with time **t** as the independent variable, together with **Eq. 15** defining **α(t)**, mathematically describe an isotropic (circular) clockwise coning motion of the CG of the bullet at the slow-mode gyroscopic precession rate $\omega_2$ and with an initial angular radius **K₀**, centered about the apparent wind direction **(0, i\*γ)**, as seen in the traditional wind axes plot for this example shown in **Fig. 3**, below.

This exercise shows how the isotropic coning motion of the CG of the bullet can be derived from the driving forces as these have been formulated above. This basic coning motion can be described as the CG of the bullet orbiting in a clockwise circular path about its mean trajectory at a cyclic rate determined by the rate of gyroscopic precession $\omega_2$ of the bullet's spin-axis.

As long as the spin-rate of a gyroscope remains nearly constant, and the overturning torque remains nearly constant, the rate $\omega_2$ of the stable slow-mode precession of the gyroscope will also remain nearly constant, and the spin-axis of the gyroscope will maintain a nearly constant angle **α** with its neutral coning axis of precession (absent any undamped $\omega_1$-rate fast-mode nutating motion).

The torque acting upon the spinning bullet is just the aeroballistic overturning moment **M**, the magnitude **{M}** of which we traditionally calculate in linear aeroballistics as:

$$\{M\} = q * S * d * Sin(\alpha) * CM\alpha \tag{19}$$

As with the aerodynamic lift force **F_L**, the overturning moment {**M**} is also an *odd function* in angle-of-attack α.

The vector rate of gyroscopic precession $\omega_2$ is related to the angular momentum vector **L** of the spinning bullet and to the overturning moment vector **M** acting on the bullet by the *vector cross-product* [6]:

$$M = \omega_2 \times L \tag{20}$$

where the angle between the two vectors $\omega_2$ and **L** is just the angle-of-attack α.

Combining these expressions for the *magnitude* of the overturning moment vector **M** yields:

$$\omega_2 * L * \sin(\alpha) = \{M\} = q*S*d*\sin(\alpha)*CM\alpha \tag{21}$$

The overturning moment coefficient **CMα** itself comprises a tabulated primary Mach-dependent function **CM₀** plus a tabular *negative* corrective coefficient **CMδ2** function multiplied by $\delta^2 = \sin^2(\alpha)$ and then algebraically summed into the coefficient **CMα** as a function of Mach-speed and angle-of-attack α.

Note that $\sin(\alpha)$ appears as a factor in the *magnitude* {**M**} on both the left and right sides of **Eq. 21** and, thus, can be divided out *for non-zero angles-of-attack* (α≠0) so that the magnitude of the coning rate $\omega_2$ (in radians per second) can be calculated from:

$$\omega_2 * L * \sin(\alpha) = q*S*d*\sin(\alpha)*CM\alpha$$

$$\omega_2 = q*S*d*CM\alpha / L \qquad (\alpha, L \neq 0) \tag{22}$$

Also notice that the pseudo-regular precession rate $\omega_2$ is *not directly dependent* on the amplitude of the cone angle α. The variation of the overturning moment coefficient **CMα** with angle-of-attack α is quite small for the small α-angles considered here. The angular momentum **L** of the spinning bullet is the product of its moment of inertia **I_x** about its spin-axis and its instantaneous vector spin-rate ω (ω = 2*π*p), so **L**≠0 for any spinning bullet.

The vector spin-rate **p** of the bullet in revolutions per second (or hertz) slows *only very gradually* in flight in accordance with an aeroballistic spin-damping coefficient.

Also note that, from Tri-Cyclic Theory, $\omega_1 + \omega_2 = (I_x/I_y)*\omega$, so the nutation rate $\omega_1$ can also be calculated readily.

## Wind Shift Effects

Whenever the coning bullet encounters a new wind **W**, its massless cone axis *can* and necessarily *does* move quickly so as to point directly into the new apparent wind **W$_A$**, controlled by the three-dimensional vector relationship:

**W$_A$** = **W** – **V**  (23)

The apparent wind **W$_A$** is just the true wind vector **W** (in an earth fixed coordinate system) translated into a coordinate system moving at velocity **V** along with the bullet.

Envision for the moment a horizontally fired, perfectly launched, spin-stabilized, ideal rifle bullet that has just emerged from the muzzle-blast cloud and has not yet begun any actual (non-zero) coning motion.

If a steady crosswind **W**, of much slower speed than **V**, is blowing horizontally from 9:00 o'clock (i.e., from left-to-right so that **W** is negative) across the trajectory, this non-coning bullet will experience an approaching apparent wind vector from just leftward of straight ahead. The small *negative* wind angle γ is given in radians *in this particular case* by:

γ = Tan$^{-1}$[W/V] ≈ W/V        [W<<V]  (24)

This angular difference γ between the **–V** direction and the apparent wind direction **W$_A$** creates a small rightward cross-track component of the downwind aerodynamic drag force **F$_D$**, of magnitude -γ*F$_D$, which in turn causes the familiar rightward horizontal wind drift of a rifle bullet fired through this constant crosswind as first formulated by Didion in 1859. As an aeroballistic angle-of-attack for our rotationally symmetric bullet, this wind angle γ will be treated as an inherently non-negative magnitude value.

As a result of this left-to-right crosswind **W**, the horizontally fired rifle bullet, with its CP ahead of its CG, immediately begins experiencing a *nose-rightward* aerodynamic overturning moment **M**, a torque vector pointing ***vertically downward*** in this case, and of magnitude given by **Eq. 19**:

{M} = q*S*d*Sin(γ)*CMγ

As a *gyroscopic reaction* to the application of this external torque **M** to the spinning bullet, the forward-pointing angular momentum vector **L** of the right-hand-spinning bullet will be just as strongly forced downward, *in the direction of the moment vector* **M**, and at a rate proportional to the magnitude of **M**, according to the vector relationship:

dL/dt = M  (25)

This *gyroscopic relationship* is just the rotational analogue of Newton's *Second Law of Motion*. Of course, the nose of the rigid bullet is pulled downward along with the bullet's angular momentum vector **L**.

## The Crosswind Aerodynamic Jump

As the coning motion is becoming established for this originally non-coning, horizontally fired bullet which is just encountering a purely horizontal left-to-right crosswind, its spin-axis direction will initially accelerate *rightward* and then *predominately downward* from its original orientation in the **+V**-direction. This is clearly shown in **Fig. 6**; a computer-generated wind axes plot from 6-DoF data provided by Bryan Litz.

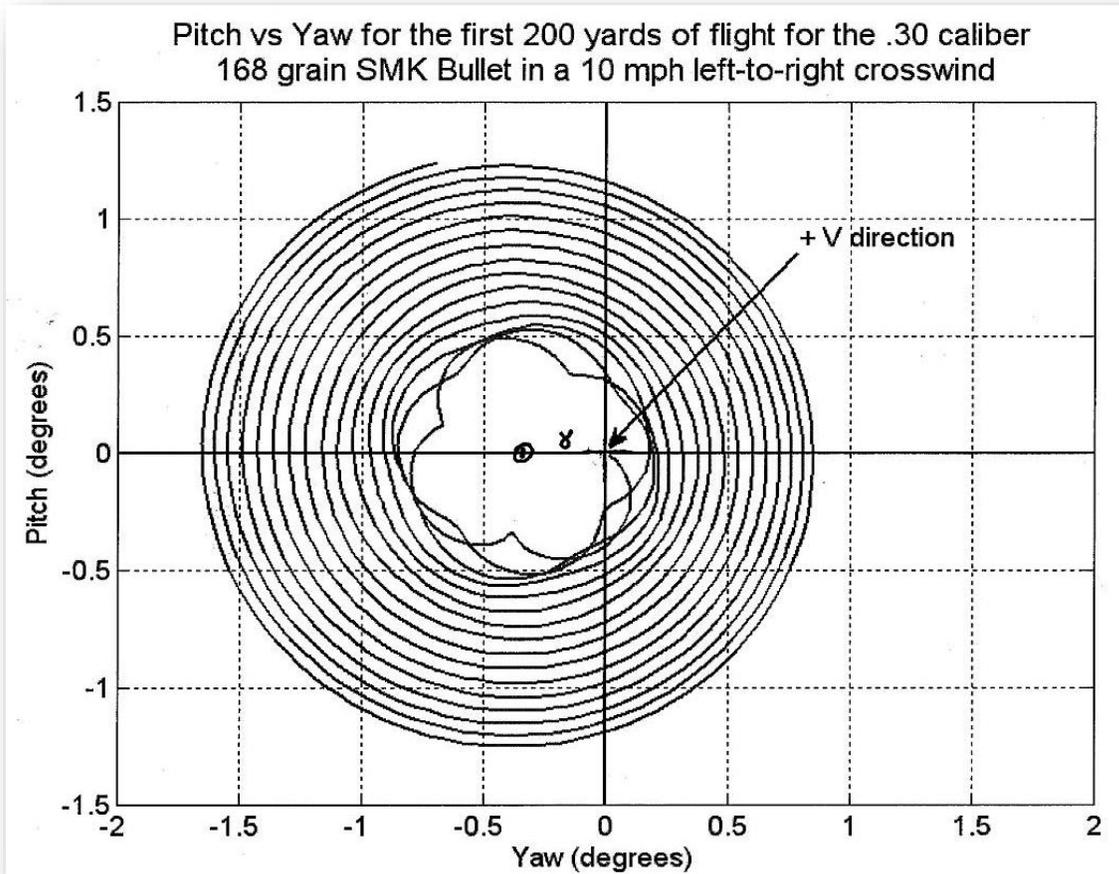

Fig. 6   Epicyclic Motion of Spin-Axis Direction (Provided by Bryan Litz)

During the coning start-up period, with the initial movement of the spin-axis being rightward in the downwind direction of the crosswind **W**, the bullet's trajectory becomes permanently deflected slightly rightward and more downward by a transient, aerodynamic impulse **J** due to its momentarily nose-right and nose-down attitude relative to the approaching windstream. After completion of the first half (180 degrees) of the slow-mode coning cycle, the *uniformly rotating* aerodynamic lift force vector $F_L$ produces no further *net* deflection of the bullet's path; i.e., the time integral of the steadily precessing lift vector approaches a limit of zero net force over successive cycles of nutation and precession (coning) motion. A practical calculation of a rifle aiming elevation correction for this vertical direction CWAJ has been separately published [15] with co-author Gustavo F. Ruiz of Patagonia Ballistics, LLC.

Actually, the suddenly encountered crosswind **W** creates a non-zero fast-mode (nutation) arm of initial angular magnitude $\delta_0$, as well as a slow-mode (precession) arm of magnitude $\gamma_0+\delta_0$. Each arm rotates clockwise for our right-hand-spinning rifle bullet, but the initial roll-orientations ($\xi_1$ and $\xi_2$) of the two epicyclic arms are *oppositely directed*, so that the slow-mode precession arm ($\gamma_0+\delta_0$) initially points rightward, downwind from the apparent wind direction (0, i*$\gamma_0$) along the **+yaw** axis, and the fast-mode nutation arm ($\delta_0$) initially points back to the left in the **-yaw** direction.

Their combined *epicyclic sum* must initially equal $\gamma_0$, the offset angle due to the initial apparent wind at time **t = 0** when the bullet has just exited the muzzle blast cloud, because the spin-axis of this perfectly launched bullet starts out pointing directly in that initial **+V** direction (the origin of the wind axes plot). This equality of the vector sum of the two epicyclic arms across any step-change in flight conditions conserves the angular momentum of the spinning bullet as physics requires.

As shown in **Fig. 6**, the fast-mode nutations damp fairly rapidly to insignificance after 5 or 6 slow-mode coning cycles for our example bullet. After completion of the first half of the first coning cycle at **$f_2$ hertz** where the lift force begins reversing its vertical-direction effect, the steadily rotating aerodynamic lift forces acting on the coning bullet subsequently tends to integrate out to zero net additional deflection of the trajectory. Certainly the rotating lift force attributable to the fast-mode (**$f_1$ hertz**) nutation itself averages to a net of zero even more quickly.

The initial epicyclic motion of the bullet's spin-axis about the apparent wind direction in wind axes coordinates can be defined in terms of the complex cone angle function **α(t)** as:

$$\alpha(t) = i*\gamma_0 + (\gamma_0+\delta_0)*(\cos\xi_2 + i*\sin\xi_2) + \delta_0*(\cos\xi_1 + i*\sin\xi_1) \qquad (26)$$

These three vector terms are, respectively; **§1** the initial and nearly constant apparent wind offset angle $\gamma_0$; **§2** the initial slow-mode, clockwise rotating coning motion arm of length $\gamma_0+\delta_0$; and, **§3** the initial fast-mode, clockwise rotating nutation arm of length $\delta_0$. These vectors are best envisioned as being summed head-to-tail in this rate-dependent sequential order. As the radii of their respective epicyclic arms in **§2** and **§3** should be inherently non-negative values, we use the absolute magnitudes of $\gamma_0$ and $\delta_0$ in formulating those two rotating vectors. In particular, the length of the slow-mode coning arm is the scalar sum of the magnitudes of its two components.

The apparent wind vector **§1** is always defined with respect to the **–V** direction. This **–V** vector points toward the origin of the "wind axes" coordinate system, but from behind the plot. $\xi_1(t)$ and $\xi_2(t)$ are the instantaneous phase angles of the clockwise rotations of the respective fast and slow epicyclic arms at any time $t$.

These phase angles are measured positive clockwise from the $+\varphi$ (pitch) axis direction. The fast arm §3 rotates clockwise at the gyroscopic nutation rate $\omega_1$, while the slow arm §2 rotates clockwise at the gyroscopic precession (or coning) rate $\omega_2$.

The angular arguments $\xi_1(t)$ and $\xi_2(t)$ start out as explained above and grow with ongoing time $t$ at their respective rotation rates, so that:

$\xi_1(t) = \omega_1{*}t - \pi/2$

$\xi_2(t) = \omega_2{*}t + \pi/2$ (27)

The spin-axis epicyclic amplitude angles $(\gamma+\delta)$ and $\delta$ also vary slowly with time $t$ according to their respective slow-mode and fast-mode *exponential damping factors*. For our somewhat *dynamically unstable* example bullet, its slow-mode damping exponent is a small positive value, and the coning angle $\alpha(t)$ slowly *increases* with ongoing time $t$.

The *vertical (pitch) component* of this epicyclic motion $\varphi(t)$ at any time $t$ during the initial relative nutation cycle, is given by the *real part* of the expression (**Eq. 26**) for $\alpha(t)$:

$\varphi(t) = Re\{\alpha(t)\} = (\gamma_0+\delta_0){*}[Cos(\omega_2{*}t + \pi/2)] + \delta_0{*}[Cos(\omega_1{*}t - \pi/2)]$

$\varphi(t) = -(\gamma_0+\delta_0){*}[Sin(\omega_2{*}t)] + \delta_0{*}[Sin(R{*}\omega_2{*}t)]$ (28)

From the initial condition of *rightward first motion*, we can set the time-derivative of this vertical component $\varphi(t)$ expression (**Eq. 28**) *initially equal to zero*, so that:

$\gamma_0 + \delta_0 = R{*}\delta_0 = 4.79{*}\delta_0$

$\delta_0 = \gamma_0/(R - 1) = \gamma_0/3.79$ (29)

Thus, the initial condition of *rightward first motion* of the spin-axis from the $+V$ direction determines the size of the nutation arm $\delta_0$ relative to the initial size of the apparent wind offset $\gamma_0$ via the initial ratio $1/(R–1)$.

From the conditions established for this flight simulation, we know that the initial magnitudes are:

$\gamma_0 = (-14.67\ fps)/(2800\ fps) = -5.24$ milliradians (30)

$\delta_0 = \gamma_0/3.79 = -1.38$ milliradians (31)

$(\gamma_0+\delta_0) = -6.62$ milliradians (in this example) (32)

The vertical-direction aerodynamic impulse $J_V$ can be calculated as the time integral of the vertical component of the perpendicular aerodynamic coning force $F_C(\alpha, t)$ over the first $n$ complete fast-mode nutation cycles relative to the moving slow-mode arm, which spans the time required to establish a steady coning motion at the precession rate $\omega_2$. The integer number $n$ of these complete relative nutation cycles available for integration of the vertical-direction impulses $J_V$ is calculated as:

$$n = \text{INT}((R-1)/4) + 1 \tag{33}$$

Note that $n = 1$ for this example. If we were numerically integrating the coning force $F_C$ over time, we would integrate from $t = 0$ to $t = T_2/2$ where the vertical component of the coning force begins to reverse itself. Here, we choose the first $n$ complete relative nutation cycles for closed-form mathematical tractability.

The number $n$ assures that we are integrating over at least the first $\pi/2$ radians (90 degrees), but no more than $\pi$ radians (180 degrees), of the first slow-mode coning motion.

The time $T_N$ of completion of the first full *relative* nutation cycle is found (with $R = 4.79$ initially in this example) from:

$$T_N = 2\pi/[\omega_1 - \omega_2] = 2\pi/[(R-1)*\omega_2] = T_2/3.79 = 4.06 \text{ milliseconds} \tag{34}$$

While the coning motion is establishing itself, the vertical (pitch) component of the total epicyclic angle-of-attack $\alpha(t)$ is increasing in magnitude from zero to a maximum amplitude of about $1.35*\gamma$ (pitch downward) from the wind-axes plot for this example.

A vertical-direction impulse $J_V$ (in pound-seconds) directly causes a vertically downward *change* in the linear momentum of the bullet, which can be well-approximated by use of a linearizing technique:

$$J_V = \int F_C[\varphi(t)]dt = n*T_N*\text{AVE}\{F_C(\varphi(t))\} \approx n*T_N*F_C[\text{AVE}\{\varphi(t)\}] \tag{35}$$

and

$$J_V = \Delta(mV_C) = m*\Delta V_C \tag{36}$$

Where the definite time integration runs from $t = 0$ to $t = n*T_N$ and $\Delta V_C$ is a negative, or downward, kick velocity in this example.

Here we are using Coning Theory to full advantage in calculating the crosswind aeroballistic jump impulse by evaluating the coning force $F_C$ perpendicular to the spin-axis of the bullet instead of simply the lift force $F_L$ attributable to this average vertical-direction angle-of-attack $\text{AVE}\{\varphi(t)\}$. Initially, at $t = 0$, the spin-axis of the bullet is perfectly aligned with the $+V$ direction of its trajectory before any non-zero coning motion has begun.

The time-average of the vertical component of the angle-of-attack $\text{AVE}\{\varphi(t)\}$ can be evaluated by the definite integration of $\varphi(t)$ from $t = n*T_N$ to $t = 0$, reversing the integration direction to account for the direction in which the crosswind is actually moving, instead of the direction from which it is approaching. Over the first $n$ complete relative nutation cycles:

$$\text{AVE}\{\varphi(t)\} = (1/(n*T_N))\int\{[(\gamma_0+\delta_0)/\omega_2]*\text{Sin}(\omega_2*t)d(\omega_2*t) + [\delta_0/(R*\omega_2)]*\text{Sin}(R*\omega_2*t)d(R*\omega_2*t)\}$$

$$\text{AVE}\{\varphi(t)\} = (\gamma_0+\delta_0)*\{((R-1)/(n*2\pi))*[1 - \text{Cos}(n*2\pi/(R-1))] + ((R-1)/R)*(\delta_0/(n*2\pi))*\{1 - \text{Cos}[(R/(R-1))*n*2\pi]\}$$

Since the angular arguments, $n*2\pi/(R-1)$ and $n*2\pi*R/(R-1)$, always differ algebraically by $n*2\pi$, their cosines are equal, and since $\delta_0 = \gamma_0/(R-1)$ from **Eq. 29**:

$$\text{AVE}\{\varphi(t)\} = \gamma_0*((R^2-1)/(n*2\pi*R))*[1 - \text{Cos}(n*2\pi/(R-1))]$$

$$\text{AVE}\{\varphi(t)\} = -4.151 \text{ milliradians} \tag{37}$$

Note from the slow-mode angular argument ($2\pi/3.79$) that the coning motion $\xi_2(n*T_N)$ has progressed through **95 degrees** to the first inward-pointing epicyclic cusp at time $n*T_N$ while the fast-mode argument $\xi_1(n*T_N)$ has progressed through **455 degrees**:

$$\xi_1(n*T_N) = 4.79*95 \text{ degrees} = 360 + 95 \text{ degrees} = 455 \text{ degrees} \tag{38}$$

Then

$$J_V \approx (n*T_N)*F_C(-4.15 \text{ mrad}) = (n*T_N)*(q*S)*\text{Sin}(-4.15 \text{ mrad})*[C_{L\alpha} + C_D]$$

$$J_V \approx (4.06 \text{ msec})*(4.82 \text{ lbs.})*(-.00415)*[2.85 + 0.32] \approx -2.58 \times 10^{-4} \text{ pound-seconds} \tag{39}$$

The permanent vertically downward angular deflection $A_{JV}$ of the trajectory of the bullet resulting from this downward cross-track kick velocity $\Delta V_C$ (from **Eq. 35**) is given by:

$$A_{JV} = \text{Tan}^{-1}[\Delta V_C/V] \approx \Delta V_C/V = J_V/(m*V) \text{ (in radians)}$$

$$A_{JV} \approx -2.58 \times 10^{-4} \text{ pound-seconds}/2.09 \text{ pound-seconds} \approx -0.123 \text{ milliradians} \tag{40}$$

Note that $m*V = 2.09$ **pound-seconds** is just the initial linear momentum of the bullet itself at the time ($t = 0$) at which it first encounters the crosswind **W**.

The permanent, one-time-per-firing, downward angular deflection $A_{JV}$ of the entire remaining trajectory produces a downward displacement on the target that varies *linearly* with *range to the target* (minus about 10 yards). At a target distance of **200 yards**, the deflected bullet strikes **0.841 inches** *below* the center of the bullseye.

This vertically downward (negative), angular deflection $A_{JV}$ of the bullet's trajectory, is one type of *aerodynamic jump*, as it has been termed by Bob McCoy of BRL, caused by encountering a purely horizontal crosswind.

McCoy [11] formulates his vertical-direction crosswind aerodynamic jump $J_A$ as:

$$J_A = - [I_x/(m*d^2)]*(C_{L\alpha}/C_{M\alpha})*(2\pi/n)*(W/V) \qquad \text{(Eq. 12.98, McCoy)}$$

$$J_A = - [0.10854]*(2.85/2.56)*(2\pi/38.96)*(14.67/2800) = -0.102 \text{ milliradians} \tag{41}$$

$J_A = 0.829*A_{Jv}$ (our value calculated here from coning theory)   (42)

Had McCoy used $(CL\alpha+CD)/CM\alpha$ as proposed herein, instead of just $(CL\alpha/CM\alpha)$ in his formulation for $J_A$, he would have found:

$J_A = -0.114$ milliradians $= 0.922*A_{Jv}$ (the coning value)   (43)

or a value **7.8 percent** smaller than our calculated $A_{Jv}$.

If the initial crosswind **W** had been blowing from right-to-left **or** if the bullet had instead been fired with left-hand spin, the angular vertical deflection of the trajectory would have been positive (upward) instead of negative (downward).

The horizontal-direction impulse $J_H$ is always in the *downwind* direction for either sense of bullet spin; i.e., first motion is horizontally downwind followed by the epicyclic motion curling either upward or downward depending upon bullet rotation sense.

We must use a different approach in evaluating the horizontal-direction impulse $J_H$ because the coning force $F_c$ reverses its horizontal direction after only **π/2 radians** (90 degrees) of epicyclic motion.

This formulation looks primarily at the slow-mode precession of the bullet's spin-axis, and incorporates the (smaller) fast-mode nutation motion only as an average increase in the coning angle $\gamma_0$. We have to stop integrating the horizontal impulse $J_H$ after only the first 90 degrees of coning motion, at time $T_2/4$, by which time the horizontal lift-force reverses and the transient horizontal impulse must have been completed.

Recall that: $T_2 = 1/f_2$ (in seconds). Then, from coning theory, the horizontal impulse $J_H$ can be calculated to be:

$J_H = -(T_2/4)*(2/\pi)^2 * \gamma_0*SQRT((R + 1)/(R – 1))*(q*S)*(CL + CD)$   (44)

This formulation incorporates a factor of $(2/\pi)*[SQRT(R^2 – 1)]/[R*Cos^{-1}(1/R)]$ to account for averaging of the **Cosine(coning phase angle)** function over its first quadrant.

Then the *rightward* horizontal angular deflection $A_{Jh}$ can be calculated from:

$A_{Jh} = J_H/(m*V)$ (in radians)

$A_{Jh} = 0.1129$ milliradians   (45)

This horizontal component of the crosswind aerodynamic jump always acts downwind and reverses with crosswind direction, but *not* with the sense of the rifling twist and the resulting spin-direction of the bullet.

However, this horizontal jump $A_{Jh}$ only applies at about 10 yards downrange. By about 20 yards downrange, the continuing coning motion of the bullet has almost completely cancelled it to **zero** net effect. *Thus, the horizontal component of this crosswind aerodynamic jump is not significant as a rifle aiming correction.*

Additionally, similar one-time transient aerodynamic jumps will occur whenever an already-coning bullet encounters any significant change in the direction **γ** of the approaching apparent wind.

We analyze a case here starting with a non-coning bullet for convenience, but also because this initial Crosswind Aerodynamic Jump (CWAJ) can be calculated as an aiming correction requiring measurement of the crosswind *only at the firing point*.

If a bullet is launched with some non-zero initial pitch and (or) yaw attitude, a similar aerodynamic jump will occur very early in its flight, as well. Unfortunately, real rifle bullets routinely suffer in-bore yaw or muzzle brake induced yaw during firing and enter the airstream with a greatly enlarged first-maximum aeroballistic yaw angle as given by Kent's Equation (Eq. 12.92, McCoy) [12].

A bullet's having a non-zero initial *yaw rate* will also produce a similar type of one-time aerodynamic jump. In any event, the rotational orientation of the one-time *jump angle deflection* of the bullet's path will always be *rotated 90 degrees clockwise* (as seen from behind the bullet) from the roll-orientation of the *initial movement* of the spin-axis of a clockwise-spinning bullet.

This 90-degree-advanced clockwise rotation for any aerodynamic jump effect is indicated in the complex plane by the initial factor of **i** (where **i$^2$ = -1**) in the BRL formulation (Eq. 12.83, McCoy) [13] for these types of aerodynamic jump.

The angular deflection of the trajectory caused by any of these types of initial aerodynamic jump effects gets established during the first half-coning cycle which take place during the first 10 yards of travel after entering the undisturbed atmosphere for most rifle bullets.

Perhaps this is one of the reasons why experienced rifle competitors pay particular attention during matches to any wind blowing across directly in front of their firing positions.

## Yaw-of-Repose and Spin-Drift Introduction

The yaw-of-repose angle $\beta_R$ is a very small, but gradually increasing, horizontally rightward, aircraft-type yaw-attitude bias or "side-slip" angle of the *coning axis* of a right-hand spinning bullet. The yaw-of-repose reverses sign and angles leftward for a left-hand spinning bullet. We discuss only right-hand spinning bullets here for clarity. We shall show that the acceleration of gravity acting upon the flat-fired bullet in free flight is the original cause of this small horizontal attitude bias. The yaw-of-repose is constrained to work only in the horizontal plane because it must be perpendicular to the local gravity field.

If the 3-dimensional mean trajectory of a rifle bullet in nearly horizontal flat firing is projected down onto a horizontal plane as shown earlier in **Fig. 5**, the rightward deviation $\beta_T$ of its tangent direction from the firing azimuth essentially defines this yaw-of-repose angle $\beta_R$ throughout the flight, except for a smaller horizontal tracking error $\varepsilon_H$ as the trajectory curves to the right *following* (lagging behind and driven by) the slightly larger yaw-of-repose angle $\beta_R$:

$$\beta_R = \beta_T + \varepsilon_H \tag{46}$$

We will formulate a good approximation for $\beta_T$ as an aid in formulating $\beta_R$ accurately.

The horizontal spin-drift **SD** which we observe in long-range shooting is due to a horizontally acting aerodynamic *lift force* attributable to the increasing yaw-of-repose attitude angle $\beta_R$ of the coning rifle bullet. We will use the principles of linear aeroballistics in formulating the yaw-of-repose and its resulting spin-drift.

Detailed analyses of PRODAS 6-DoF simulation runs show that in flat firing the magnitude of the spin-drift **SD**, in any given simulated firing after the first 150 yards or so of flight, is equal to some invariant scale factor **ScF** of about **1** to **2 percent**, more or less, times the bullet's *drop from the projected bore axis*:

$$SD(t) = -ScF*DROP(t) \tag{47}$$

In other words, the horizontal spin-drift trajectory looks just like a small fraction **ScF** of the vertical trajectory rotated 90 degrees about the axis of the bore with each curvature ultimately caused by the same gravitational effect.

We must formulate the scale factor **ScF** so that it can be evaluated accurately for any given bullet type and firing conditions. Then using **Eq. 47**, we need only an accurate determination of the bullet's drop from the bore axis at the target distance to calculate an accurate horizontal spin-drift across the face of that long-range target. Existing point-mass (3-DoF) trajectory programs specialize in the accurate calculation of this bullet drop at the target distance in any firing conditions. A procedure for calculating the scale factor **ScF** is included in the separately published referenced paper **[16]**.

## The Horizontal Tangent Angle

The instantaneous tangent to the horizontal-plane projection of the mean trajectory forms the angle $\beta_T(t)$ to an **X**-axis in that horizontal plane which defines the launch azimuth of the fired bullet. The *mean trajectory* of the bullet is the 3-dimensional path which would be followed by the CG of the bullet if it were not coning about that mean trajectory. This horizontal tangent angle $\beta_T(t)$ is always defined by the horizontal projection of the bullet's mean velocity vector $V(t)$, but these mean velocity components are not calculated in our available PRODAS reports. The rifle bullet's reported instantaneous velocity components are modulated by the helical coning motion of the CG of the bullet in flight.

As the bullet drifts horizontally due solely to spin-drift $SD(t)$, the intersection point of the tangent to the bullet's mean trajectory with the **X**-axis moves forward in the **+X** direction, but at a slower velocity than the forward velocity of the bullet itself. If we assume a continually increasing curvature of the horizontal trajectory so that this velocity ratio varies exponentially with range $X(t)$, we can estimate $\beta_T$, the dominant portion of $\beta_R$, as:

$$\beta_T \approx \mathrm{TAN}(\beta_T) = SD(t)/\{X(t)*0.825*\exp[-0.925*X(t)/X(max)]\} \qquad (48)$$

This hand-fitted estimator function agrees quite well with $\beta_T(t)$ angular values extracted from available trajectories generated by PRODAS 6-DoF simulations for our particular long-range bullet by ratioing an extracted rightward $V_R(t)$ to $V(t)$ for each millisecond of the PRODAS trajectory reports. The $V_R(t)$ data is extracted by applying a smoothing difference operator to the PRODAS "no wind, no Coriolis" drift data converted into linear distance units. Comparing the two functions for each millisecond over the **1.6923-second** simulated flight time yields a mean difference of **1.12 micro-radians** with a population standard deviation of **0.0514 milliradians**. The damped "epicyclic swerve" modulation of the spin-drift displacement $SD(t)$ is greatly reduced in significance as the magnitude of $SD(t)$ grows larger over time. Extraction of the small rightward horizontal velocity $V_R$ from the trajectory drift data is complicated by the superimposed epicyclic swerving of the CG of the bullet which accounts for most or the variance between these two functions.

We need to formulate an approximation for $\beta_T(t)$ so that it can be used as a reasonableness check on formulations of $\beta_R(t)$ which is not itself reported by PRODAS. We will need an accurate formulation for $\beta_R(t)$ in order to calculate the scale factor **ScF** and thence the spin-drift $SD(t)$ for other rifle bullets without relying upon 6-DoF simulation data. This spin-drift calculation is detailed in a separate paper **[16]** jointly published with co-author, Gustavo F. Ruiz.

## PRODAS Simulated Bullet Flight Data

In this study we will use as our example bullet the 30-caliber 175.16-grain bullet as loaded in the US Army's M118LR Special Ball (7.62x51 mm NATO) long-range sniper and match ammunition. We do this because we have several PRODAS 6-DoF simulation runs on hand for this 7.62 mm NATO ammunition, reporting the linear ballistic results (including spin-drift) for each millisecond of its **1.6923-second** total simulated flight time to **1000 yards**.

The simulated firing conditions are 1) flat firing, 2) standard sea-level ICAO atmosphere, 3) no wind, 4) no Coliolis effect calculated, 5) muzzle velocity of **2600.07 feet per second**, and 6) barrel twist is right-handed at **11.5 inches per turn**. The "no-wind" and "no-Coliolis" conditions assure that the rightward spin-drift **SD** is the only secular horizontal "bullet drift" being calculated by PRODAS. However, the PRODAS reported drift data necessarily includes the oscillating horizontal component of the bullet's helical coning motion about its mean trajectory throughout its simulated flight. We also have PRODAS runs available for this same bullet fired through constant left and right 10 MPH crosswinds as well as left-hand twist runs in each of the three constant wind conditions.

## Yaw-of-Repose

We will show that in flat firing the continual downward arc of the flight path angle **Φ** due to gravity causes repeated rightward differential aerodynamic torque impulses **ΔM** centered about the extreme top-dead-center (TDC) and bottom-dead-center (BDC) positions of the CG of the bullet in its coning motion. These double-rate yaw attitude-changing torque impulses cause the forward-pointing angular momentum vector **L** of the right-hand spinning bullet to shift evermore rightward during its flight.

Ballisticians term this accumulating yaw-attitude bias the "yaw-of-repose" angle $\beta_R$ of the flying bullet and classically formulate it from calculus as [**Eq.10.83** in McCoy's MEB]:

$$\beta_R = P*G/M \tag{49}$$

in terms of the classic aeroballistic auxiliary parameters:

$$P = (I_x/I_y)*p*d/V = (\omega_1 + \omega_2)*d/V \tag{50}$$

$$G = g*d*\cos\varphi/V^2 \approx g*d/V^2 \tag{51}$$

$$M = (m*d^2/I_y)*[\rho*S*d/(2*m)]*C_{M\alpha} = (\omega_1 + \omega_2)*\omega_2*d^2/V^2 \tag{52}$$

after converting each classic auxiliary parameter from dimensionless arc-length-rates into the time-rate units used here for our analyses of flat-firing a spin-stabilized rifle bullet.

The change-of-variables in **Eq. 50** uses the relation from Tri-Cyclic Theory that:

$$(I_x/I_y)*p = (I_x/I_y)*\omega = \omega_1 + \omega_2 \tag{53}$$

McCoy defines the canonical spin-rate **p** of the bullet as used here to be a circular frequency given in radians per second. The bullet's spin-rate **p** is sometimes given elsewhere in aeroballistics in units of revolutions per second (or hertz) as used earlier in this paper, or is sometimes given in radians per foot of bullet travel, or even in radians per caliber of bullet travel. To avoid this confusion we use the more conventional symbol ω here for the circular frequency of the spin-rate of the bullet given in radians per second.

The change in **Eq. 52** uses the fundamental magnitude relationship from Coning Theory that:

$$(\rho * S * V^2 / 2) * d * CM\alpha = L * \omega_2 = (Ix * \omega) * \omega_2 \tag{54}$$

as well as the Tri-Cyclic relation in **Eq. 53** again.

With these changes of variables, the classic formulation for the yaw-of-repose angle $\beta_R$ reduces to:

$$\beta_R = g / [\omega_2(t) * V(t)] = g / [2\pi * f_2(t) * V(t)] \tag{55}$$

While this formulation for $\beta_R$ is classic, it *does not inherently yield* **zero** *at* **t = 0**, and it is about a factor of **π** too small at long ranges when compared to $\beta_T$ as formulated above.

Let us say the flight path angle Φ of the bullet's trajectory changes downward by a small decrement ΔΦ due solely to the pull of gravity during one-half of the period $T_2$ of a particular coning cycle. This would be the case for a vacuum trajectory suffering no air drag. As a continuous variable in flight time **t**, this angular decrement **ΔΦ(t) = 0.0** at **t = 0** by definition.

In flat firing, the small decrement ΔΦ in the nearly horizontal flight path angle Φ(t) during the time interval $T_2/2$ of a particular half-coning cycle can be expressed as:

$$\Delta\Phi(t) \approx TAN(\Delta\Phi) = -(g * T_2) / [2 * V(t)] = -g / [2 * f_2(t) * V(t)] = -\pi * g / [\omega_2(t) * V(t)] \tag{56}$$

where $f_2(t)$ is the instantaneous coning rate, or gyroscopic precession rate, of the bullet in revolutions per second, or hertz.

Comparing our version of the classic formulation for the steady-state yaw-of-repose $\beta_R(t)$ in **Eq. 55** with the change in flight path angle ΔΦ(t) *per half-coning cycle* above, we note that:

$$\beta_R(t) = (-1/\pi) * \Delta\Phi(t) \tag{57}$$

Thus our formulation in **Eq. 56** above for **ΔΦ(t)**, the change in flight path angle Φ per half-coning cycle $T_2/2$ which *does* inherently equal **zero** at **t = 0**, actually looks like a better formulation for $\beta_R(t)$ than does the classic form.

We will investigate the aerodynamic and gyroscopic causes of $\beta_R(t)$ so that we can formulate its value at any time **t** during the flight of any rifle bullet. With ongoing time-of-flight, the downward arcing of the trajectory due to gravity continually causes the airstream to approach the flying bullet from below its spin-axis direction.

At each extreme location, TDC and BDC, the coning bullet experiences a peak rate of differential change in its aerodynamic overturning moment vector **M** due to this differential change **ΔΦ** in its vertical-direction aerodynamic angle-of-attack. Each of these two differential torque impulse vectors **ΔM** points *horizontally rightward* as seen from behind the right-hand spinning bullet. Here these differential *torque impulse* vectors **ΔM** are to be evaluated by integrating the differential torque over each half of the coning period **T₂**, giving them units of torque multiplied by time which can be equivalenced to the units of angular momentum.

Owing to the increased aerodynamic angle-of-attack of the apparent wind experienced by the bullet at its BDC position, the differential torque impulse **ΔM** at BDC is *inherently positive rightward*, temporarily increasing the overturning moment **M** acting upon the bullet at this BDC location in its coning motion. Recall that in Coning Theory the spin-axis of the bullet is pointing maximally *upward* when the CG of the bullet is at its BDC position in any coning cycle.

As formulated in linear aeroballistics, the instantaneous magnitude {**M**} of the overturning moment **M** at time **t** is:

{M} = q*S*d*Sin[α(t)]*CMα          (58)

where

**q** = (ρ/2)*V² = Dynamic Pressure in lbf/square foot.

**ρ** = Density of the atmosphere = **0.0764742 lbm/cubic foot** for the standard sea-level ICAO atmosphere used here. This value of the density **ρ** must be divided by the acceleration of gravity **g = 32.174 feet per second per second** to convert its units into proper density units, mass (**slugs**) per cubic foot.

**V** = Airspeed of the bullet in feet/second.

**S** = Reference (frontal) area of the bullet at the base of its ogive in square feet = (π/4)*d².

**d** = Diameter of the bullet in feet.

**α(t)** = Coning angle of the bullet in radians at any time **t** during its flight.

**CMα** = Dimensionless overturning moment coefficient in linear aeroballistics theory.

As the flat-firing trajectory of the coning bullet, flying essentially horizontally near the **X**-axis (with **Φ ≈ 0.0** and with its coning axis aligned into the approaching windstream) arcs downward due to gravity, the aerodynamic angle-of-attack **α(t)** increases by the magnitude of **ΔΦ** at its BDC location in this coning cycle. The **cosines** of the coning angle **α(t)<5.7 degrees**, the flight path angle **Φ**, and the small change in flight path angle **ΔΦ** all remain essentially equal **1.00**. From trigonometry, the peak magnitude $\{\Delta M\}_{PEAK}$ of this differential overturning torque $\Delta M$ with the bullet at its BDC location can be expressed as:

$SIN(\alpha + \Delta\Phi) = SIN(\alpha)*COS(\Delta\Phi) + COS(\alpha)*SIN(\Delta\Phi) \approx SIN(\alpha) + SIN(\Delta\Phi)$

$M + \{\Delta M\}_{PEAK} = q*S*d*SIN(\alpha + \Delta\Phi)*CM\alpha \approx M + q*S*d*SIN(\Delta\Phi)*CM\alpha$

$\{\Delta M\}_{PEAK} = q*S*d*SIN(\Delta\Phi)*CM\alpha$ (59)

This expression can also be well approximated as:

$\{\Delta M\}_{PEAK} = q*S*d*(\Delta\Phi)*CM\alpha$ (60)

The instantaneous vertical-direction aerodynamic angle-of-attack is actually the vector sum of three small angles in complex wind-axes coordinates (ignoring the fast-mode $\omega_1$ motion):

1. Vertical component of the slow-mode coning angle, $\alpha(t)*COS(\omega_2*t + \xi_0)$
2. Downward change in flight path angle $\Delta\Phi$, and
3. Very small vertical-direction tracking error angle $\varepsilon_V$ (upward in wind-axes plots). This vertical-direction tracking error angle $\varepsilon_V$ is termed the "pitch-of-repose" by McCoy.

The primary overturning moment **M** is due to (1) the coning angle-of-attack $\alpha(t)$. This rotating torque vector **M** produces the slow-mode circular coning rate $\omega_2(t)$ of the bullet as a gyroscopic precession of the bullet's spin-axis. Examination of several different PRODAS runs shows that even for a *dynamically stable* bullet with any early coning motion fully damped down, $\alpha(t)$ always exceeds $\Delta\Phi$ by some margin all the way to maximum supersonic range and beyond.

From Coning Theory, the vertical component of the complex coning angle $\alpha(t)$ is the "pitch angle" $\varphi(t)$ given by the *real part* of the complex $\alpha(t)$, again neglecting the fast-mode motion:

$\varphi(t) = Re[\alpha(t)] = \{\alpha(t)\}*COS(\omega_2*t + \xi_0)$ (61)

Whenever $\alpha(t) >> \Delta\Phi$, only the portion of $\varphi(t)$ equal in magnitude to $\Delta\Phi$ produces the overturning moment impulse $\Delta M$ which drives the spin-axis of the bullet rightward giving rise to the bullet's yaw-of-repose angle $\beta_R$, and the overturning moment impulses at BDC and TDC can be assumed to have the same forms. The excess of $\varphi(t)$ over $\Delta\Phi$ goes toward enlarging the coning angle $\alpha(t)$, counteracting any frictional aerodynamic damping of that slow-mode coning motion of the bullet.

The instantaneous differential overturning moment $\{\Delta M\}$ is then due to the vertical-direction differential aerodynamic angle-of-attack $\Delta\Phi(t)*COS(\omega_2*t + \xi_0)$. This modulation at the coning-rate $\omega_2$ looks like a full-wave-rectified sine wave over each coning cycle. The average over each quarter wave is just $2/\pi$ times the peak value. The average value of $\Delta\Phi$ itself over each half-coning cycle is just $\Delta\Phi/2$ because the flight path angle $\Phi$ varies almost linearly over the small interval $T_2/2$. Averaged over the top or bottom one-half of a coning cycle, the effective angle-of-attack is then $(2/\pi)*(\Delta\Phi/2) = \Delta\Phi/\pi$.

The vector sum of (2) $\Delta\Phi$ and (3) $\varepsilon_V$ varies only gradually with ongoing time-of-flight $t$. The *magnitudes* of these two small angles *sum* to a vertical-direction aerodynamic angle-of-attack which drives the coning-axis direction *continually downward* according to Coning Theory, tracking (but lagging behind) the downward-curving trajectory.

The time-integrated *torque impulse* $\Delta M$ centered at TDC or BDC must equal the differential torque due to the time-average $\Delta\Phi/\pi$ aerodynamic angle-of-attack multiplied by the total time interval $T_2/2$ for each half-coning cycle. The period $T_2$ of the coning motion increases gradually as the coning rate $\omega_2$ slows throughout the flight.

The effective differential torque impulse $\Delta M$ integrated over a particular *half-coning cycle* thus becomes:

$$\Delta M = (T_2/2)*q*S*d*(\Delta\Phi/\pi)*CM\alpha \tag{62}$$

Substituting the unsigned *magnitude* of the first expression for $\Delta\Phi$ from **Eq. 56** yields:

$$\Delta M = (1/\pi)*g*(T_2/2)^2 *q*S*d*CM\alpha/V(t) \tag{63}$$

This differential torque impulse $\Delta M$ has units of lbf-feet-seconds which are equivalent to lbm-feet squared per second, a proper set of units for angular momentum.

While the overturning moment vector **M** itself points *leftward* at the TDC position of the bullet, the differential torque impulse vector $\Delta M$ is *inherently negative* due to the *reduced* aerodynamic angle-of-attack experienced by the coning bullet at that upper location, and so the differential torque impulse vector $\Delta M$ itself also points *positive rightward* at TDC. Thus, the alternating TDC and BDC differential torque impulses are *mutually reinforcing* throughout the bullet's flight.

The right-hand spinning bullet alters its pointing direction *rightward* in gyroscopic reaction to each of these two differential torque impulses $\Delta M$ during each coning cycle. But it does so in an unusual way.

When a constant-magnitude, rotating overturning moment vector **M** is applied to a spinning gyroscope, its spin-axis direction soon begins moving in precession and nutation in reaction to that steadily rotating torque vector. However, the first motion of its spin-axis is always in the direction of the eccentric force producing the overturning moment **M** while those epicyclic motions are getting started. For the spinning bullet, the eccentric force is the total aerodynamic force **F** acting through the aerodynamic center-of-pressure CP at any instant during its flight. For spin-stabilized rifle bullets, the CP is nearly always located forward of the CG along the spin-axis of the bullet.

In response to each small torque *impulse* **ΔM**, the spin-axis of our bullet moves *initially rightward*, but each impulse ceases before any *downward* movement of the spin-axis can become established. When the torque impulse vector **ΔM** is expressed in the same units as the angular momentum vector **L** of the spinning bullet, having physical dimensions of mass times length squared over time, their *direct vector sum* defines the resulting angular momentum **L** of the spinning bullet after the torque impulse has been applied.

For a right-hand spinning bullet the angular momentum vector **L** points forward along its spin-axis. Here, since **ΔM** is always acting perpendicularly to **L**, the *direction* of the angular momentum vector **L** is shifted rightward by an incremental angular amount (in radians) which we term **Δβ$_R$**, but its *magnitude* remains unchanged. The nose of the right-hand spinning bullet always points in the direction of its angular momentum vector **L**.

The incremental increase **Δβ$_R$** in the yaw-of-repose angle **β$_R$** during each *half coning cycle* is thus:

$$\Delta\beta_R \approx TAN(\Delta\beta_R) = \{\Delta M\}/\{L\} = (1/\pi)*g*(T_2/2)^2*q*S*d*CM\alpha/[L*V(t)] \tag{64}$$

Recalling **Eq. 22** from earlier in this paper, we note that the right-hand side of **Eq. 63** contains the fundamental expression from Coning Theory for determining the *magnitude* of the circular coning rate $\omega_2(t)=2\pi*f_2(t)$ for a spin-stabilized bullet coning at non-zero angles of attack **α**:

$$\omega_2 = q*S*d*CM\alpha/L \qquad (\alpha, L \neq 0)$$

Due to the acceleration of gravity, the coning angle **α(t)** cannot be **zero** in flat firing except perhaps very briefly at **t = 0**, where this magnitude relationship still holds true. The angular momentum **L** is greater than zero for any spin-stabilized rifle bullet.

After this change of variables,

$$\Delta\beta_R = (1/\pi)*g*(T_2^2)*\omega_2(t)/[4*V(t)] \tag{65}$$

This change of variables is critically important in formulating an analytical calculation of **β$_R$** because it simultaneously eliminates from the formulation both the overturning moment coefficient **CMα** and the angular momentum **L** of the bullet, each of which is difficult to calculate for a new bullet. The coning rate **ω$_2$(t)** is more readily obtainable from Tri-Cyclic Theory.

Also recall that by definition $T_2^2 = 1/(f_2)^2 = 4\pi^2/\omega_2^2$. After this substitution we have:

$$\Delta\beta_R = \pi*g*/[\omega_2(t)*V(t)] = g/[2*f_2(t)*V(t)] = g*T_2/[2*V(t)] = -\Delta\Phi \tag{66}$$

While this expression is dimensionless, the increment **Δβ$_R$** in the yaw-of-repose angle for each half-coning-cycle **T$_2$/2** is calculated here in radians. The proper sign depends upon coordinate system conventions and the sense of the bullet's spin-rate.

43 / 59    A Coning Theory of Bullet Motions – Boatright – rev. March/2018

Since the instantaneous aerodynamic lift-force driving the spin-drift **SD** of the bullet horizontally is linearly proportional to the aerodynamic angle-of-attack (for the very small yaw-of-repose angle $\beta_R$), the linear dependence of $\Delta\beta_R$ upon $\Delta\Phi$ shown in **Eq. 66** explains the remarkable similarity in shape of the horizontal-plane and vertical-plane projections of the bullet's "no wind, no Coriolis" mean trajectory in 3-space.

When $\alpha(t) \gtrsim \delta \approx \Delta\Phi$, as in most "constant wind" 6-DoF simulations, the average torque impulses $\Delta M$ are no longer equal at BDC and TDC. In fact, $\Delta M(BDC) \gtrsim \Delta M(TDC)$, and their combined average effect would be slightly smaller (by about 5 percent) than these estimates of a maximal yaw-of-repose.

The yaw-of-repose angle $\beta_R(t)$ can be found by *summing* the increments $\Delta\beta_R$ divided by $T_2/2$ for each *half coning cycle* which has occurred from **t = 0** to time **t**, starting with $\beta_R(0)$ equal **zero**.

Using the data from "no wind, no Coriolis" PRODAS reports, yields $\beta_R$(**1.430 sec**) = **0.67208 milliradians**, which exceeds our fitted value of $\beta_T$(1.430 sec) = 0.61019 mrad by **10.14 percent**. We term this difference, $\beta_R - \beta_T$, the horizontal tracking error angle $\varepsilon_H$. We are comparing these angles here at **t = 1.430 seconds** where this bullet has slowed to Mach 1.20 or **1340 feet per second** at **888.5 yards** downrange.

Using the PRODAS-calculated velocity and coning-rate data, our adjusted version of the classic formulation of the yaw-of-repose yields $\beta_R$(**1.430 sec**) = $\pi \ast P \ast G/M$ = **0.70633 milliradians**, which exceeds our fitted value of $\beta_T$(**1.430 sec**) = **0.61019 mrad** by **15.76 percent** for the tracking error $\varepsilon_H$. We believe this adjusted classic formulation for $\beta_R$ better matches the case for a significantly coning bullet than for this minimally coning trajectory.

## Analysis of the Spin-Drift

The horizontally rightward spin-drift **SD(t)** of the trajectory is caused by a net horizontal aerodynamic lift-force attributable to this small, but ever increasing, horizontally rightward yaw-of-repose angular bias $\beta_R(t)$ in the yaw-attitude of the coning-axis of the spinning bullet. The pointing direction of the bullet's coning axis quickly tracks each of these small changes in the approaching apparent wind direction within one half of a coning cycle, just as with any other type of wind change.

As the horizontal-plane projection of the bullet's mean trajectory gradually accelerates rightward with this spin-drift **SD(t)**, its tangent **+V** direction defining the origin of wind-axes plots should also properly drift slowly rightward *following* (lagging) the increasing yaw-of-repose attitude angle $\beta_R(t)$ of the bullet. We formulated this tangent angle $\beta_T(t)$ earlier. Logically, only the horizontal tracking error angle $\varepsilon_H(t) = \beta_R(t) - \beta_T(t)$ should appear in these wind-axes plots.

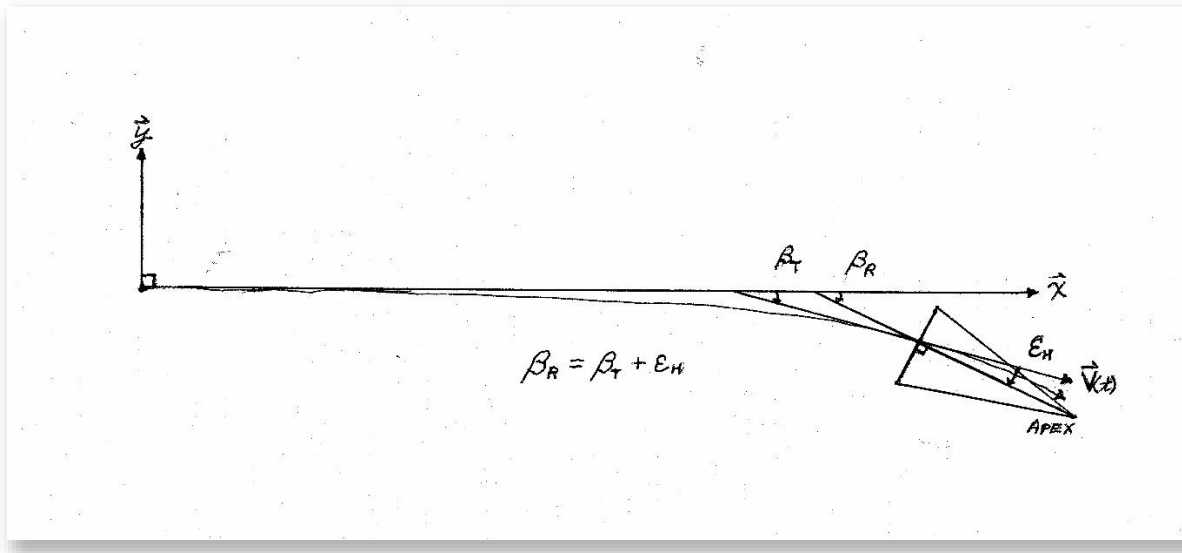

### Fig. 5   Spin Drift in Plan View for Right-Hand Spin Direction (No Wind)

In formulating the effective net (time-averaged) aerodynamic lift-force accelerating the CG of the coning bullet rightward, we must consider the coning modulation of the aerodynamic effect as the CG of the bullet moves throughout its circular coning cycle. Here the modulation is horizontally left-to-right, and the effect being modulated is aerodynamic lift.

However, for the uniformly coning rifle bullet, analysis of the modulation of this lift force can be greatly simplified by making use of Coning Theory. We can express the average effective aerodynamic lift-force on the coning bullet arising from the yaw-of-repose angle $β_R(t)$ as if the bullet were *not* coning but simply flying with the spin-axis always aligned with the attitude of its coning-axis. After all, it is the attitude of that coning-axis which defines the yaw-of-repose angle $β_R(t)$.

The magnitude of the small net rightward aerodynamic lift force $\{F_L\}_R$ attributable to the rightward yaw attitude bias $β_R(t)$ of the coning axis is given in linear aeroballistics as:

$\{F_L\}_R = q * S * CL_β * SIN[β_R(t)]$

or

$\{F_L\}_R ≈ q(t) * S * CL_β(t) * β_R(t)$ \hfill (67)

Here the coefficient of lift $CL_β(t)$ is evaluated for the very small angle-of-attack $β_R(t)$. However, the lift coefficient $CL_β(t)$ still varies with the Mach-speed of the slowing bullet. The dynamic pressure $q(t)$ also reduces with the square of $V(t)$ as the bullet slows.

This small rightward horizontal force $\{F_L\}_R$ acting on a bullet of mass **m** for one half the period $T_2$ of each coning cycle produces a rightward horizontal bullet velocity increment $\Delta V_R$ given here in **feet per second per half-coning cycle** as:

$$\Delta V_R = \{F_L\}_R * T_2/(2*m) = \{F_L\}_R/[2*m*f_2(t)] = (\pi/m)*\{F_L\}_R/\omega_2(t) \quad (68)$$

where **m** is the mass of the bullet expressed in **slugs**. Here, m = 175.16/(7000*g) = **0.00077774 slugs**. We are using g = **32.174 feet per second per second** for the standard effective acceleration of gravity on the surface of our rotating earth.

These rightward velocity increments $\Delta V_R$ accumulate (sum) from **zero** at **t = 0** for each *half coning cycle* which occurs from launch to time **t** to form the horizontally rightward velocity $V_R(t)$ of the CG of the bullet which is caused aerodynamically by the yaw-of-repose angle $\beta_R(t)$.

The incremental rightward horizontal spin-drift of the bullet, $\Delta SD$ in feet, during one particular half-coning cycle $T_2/2$ is then:

$$\Delta SD(t) = V_R(t)*T_2/2 = V_R(t)/[2*f_2(t)] = \pi*V_R(t)/\omega_2(t) \quad (69)$$

The horizontal spin-drift $SD(t)$ at time **t** is then found by *summing* these incremental displacements $\Delta SD(t)$ for each *half coning cycle* starting with **zero** at **t = 0**. Our subject bullet experiences **87 half-coning cycles** during its flight to **1000 yards**.

Numerical integration of $\Delta SD(t)$ using PRODAS data for each millisecond of the simulated "no wind" flight yields **SD(1.6923 sec) = 9.7019 inches**. PRODAS itself calculates a total drift of **9.5407 inches** at **1000 yards**. The PRODAS drift includes the horizontal component of the minimal coning motion of the spinning bullet. This level of agreement verifies our aeroballistic analysis of the causes of spin-drift.

## The Coning Angle

The coning angle $\alpha(t)$ is a *free variable* in Coning Theory as is its associated coning radius $r(t)$ which is defined as $D*\sin[\alpha(t)]$. However, the coning angle $\alpha(t)$ can be determined in general only by examining the wind-axes plots produced by 6-DoF digital simulations calculating the pitch and yaw pointing directions of the bullet's spin-axis as well as its position, velocity, and spin-rate throughout its flight. An example wind axes plot is shown in **Fig. 3** above. The coning angle $\alpha(t)$ of the bullet at time **t** depends upon the entire history of every change in flight conditions which has been encountered by the flying bullet from launch to time **t**.

The coning angle $\alpha(t)$ is *defined* to be the instantaneous angular magnitude of the slow-mode epicyclic arm rotating at the gyroscopic precession rate $\omega_2$ about the eye of the apparent wind. For a bullet just launched into a crosswind with zero yaw and yaw-rate, its initial cone angle $\alpha_0$ is equal to the sum of the magnitudes of $\gamma_0$ and $\delta_0$. Any non-zero initial aeroballistic yaw would position the initial direction of the bullet's spin-axis away from the origin of the wind axes (**+V** direction). Any significantly non-zero initial yaw-rate would likely control the direction of first movement of the spin-axis. Keep in mind that $\alpha(t)$ and $\delta(t)$ each have separate exponential damping factors in linear aeroballistics modeling.

Whenever the coning bullet experiences a step-change in the direction of the eye of the apparent wind $\gamma$, that angular step-change becomes the magnitude of a new fast-mode (nutating) epicyclic arm $\delta_N$ and the magnitude of the coning angle $\alpha$ *increases* by this same amount as the scalar sum of these magnitudes:

$$\{\alpha_N\} = \{\alpha_{N-1}\} + \{\delta_N\} \tag{70}$$

This scalar expression indicates that the magnitude of the coning angle $\alpha(t)$ can only *increase* as the bullet experiences any changes in flight conditions throughout its flight.

In accordance with **Eq. 26** above and the physical principle of *conservation of angular momentum*, the epicyclic vector sum $\alpha_N$ after the step-change $\delta_N$ must equal the vector sum just before the step-change $\alpha_{N-1}$. This requirement *fixes* the initial phase angle of the new fast-mode arm $\delta_N$ such that, as a *vector sum*:

$$\alpha_N = (\alpha_{N-1} + \delta_N) - \delta_N = \alpha_{N-1} \tag{71}$$

The minus sign $-\delta_N$ indicates that the initial fast-mode nutation phasing is **180 degrees opposite** from the direction of the step-change $\delta_N$ in the wind vector. One could envision the vector in parentheses moving the tail of the rotating slow-mode arm vector by the step-change wind vector $\delta_N$, shifting the coning axis so that it now points into the eye of the new apparent wind. Simultaneously the vector $-\delta_N$ is applied at the head end of the slow-mode arm so that the spin-axis direction remains unchanged in the wind axes plot across this step-change as required for *conservation of angular momentum* for the spinning bullet. The slow-mode arm of the coning motion steadies on the new cone angle $\alpha_{N-1} + \delta_N$ within another half coning cycle. The initial crosswind aerodynamic jump described above behaves in this manner if the initial coning angle α $_{N-1}$ is understood equal to the magnitude of γ$_0$.

These step-changes in the apparent wind vector δ$_N$ might be tiny and frequently occurring, or they might be very large and sudden one-time wind events. From *conservation of angular momentum*, the motions of the spin-axis pointing direction in the wind axes plots must always be *continuous*, even in response to a step-change in flight conditions.

While not explicitly part of Coning Theory, these relationships are presented here in detailed explanation of what we have long been seeing in 6-DoF wind axes plots.

In linear aeroballistic theory, each epicyclic arm is subject to its own exponential damping as time goes on. At any instant during the flight of a real bullet, its coning angle α(t) is determined by its complete history of flight disturbances experienced since launch and by the flight-time elapsed since the latest such disturbance.

## Energy Considerations

The ability to deliver bullets retaining sufficient kinetic energy to distant targets is crucial to riflemen. This deliverable kinetic energy **E** is calculated as

$$E(v) = (m/2)*v^2 \tag{72}$$

where          v = The ground speed of impact of the bullet.

Let us say that at any time **t** during flight, the loss in kinetic energy **ΔE** due to air drag over the small time interval **Δt** is governed by the Equations of Motion of a bullet in ballistic flight as:

$$E(t + \Delta t) = E(t) - \Delta E(\Delta t) \tag{73}$$

and

$$\Delta E(\Delta t) = F_D * \Delta s = F_D * V(t) * \Delta t \tag{74}$$

where **F$_D$** is the total aerodynamic force of drag retarding the forward motion of the bullet and **Δs** is the path length (a distance in feet) travelled by the bullet during the small time-interval **Δt**. Note that a force acting over a distance does work, which is energy. Here **V(t)** is the airspeed of the bullet at time **t**.

In particular, we are interested in the loss in kinetic energy **ΔE** during one half of any single coning cycle where

$$\Delta t = (2\pi/\omega_2)/2 = 1/(2*f_2) = T_2/2 \text{ seconds}. \tag{75}$$

In linear aeroballistics theory, the drag force **F$_D$** is accurately modelled as

$$F_D = q*S*(CD_0 + \delta^2 *CD_\alpha) \tag{76}$$

where $\delta = \text{Sin}(\alpha)$.

$CD_0$ is the coefficient of minimum drag for exactly nose-forward "zero-yaw" aerodynamic flight at a given airspeed (expressed as a Mach Number), and $CD_\alpha$ is the yaw-drag coefficient at that same airspeed for an aerodynamic angle-of-attack $\alpha$. Here this angle-of-attack $\alpha$ is the same as the coning angle $\alpha$.

Now the expression for the total kinetic energy loss during any particular half coning cycle becomes

$$\Delta E(T_2) = q*S*(CD_0 + \alpha^2 *CD_\alpha)*V*T_2/2 \tag{77}$$

For the very small coning angles-of-attack we are concerned with here, we can set

$$\text{Cos}(\alpha) = 1.00$$

and $\delta = \text{Sin}(\alpha) = \alpha(\text{in radians}) =$ Average value of $\alpha$ over the half coning cycle.

The kinetic energy loss due to yaw-drag $\Delta E_\alpha$ over this half coning cycle is then

$$\Delta E_\alpha(T_2/2) = q*S*V*CD_\alpha*(T_2/2)*\alpha^2 \tag{78}$$

We can also formulate the kinetic energy $E_C$ of the orbital coning motion itself as

$$E_C = (m/2)*(r*\omega_2)^2 = (m/2)*(D*\text{Sin}\alpha*\omega_2)^2 \tag{79}$$

$$E_C = (m/2)*(D*\omega_2)^2 *\alpha^2 \tag{80}$$

where $r$ is the coning radius of the CG of the bullet orbiting around a "mean CG" location moving along the "mean trajectory" of the bullet and $D$ is the slowly varying coning distance of the CG of the bullet from its coning apex, each given in feet, so that $r = D*\text{Sin}(\alpha)$.

Now, as the coning angle $\alpha$ decreases, due to frictional damping, from its initial value $\alpha_0$ to its final value $\alpha_1$ at the completion of this half coning cycle, the change $\Delta E_C$ in orbital coning energy can be written as

$$\Delta E_C = (m/2)*(D*\omega_2)^2 *(\alpha_0^2 - \alpha_1^2)$$
$$\Delta E_C = [m*(D*\omega_2)^2]*[(\alpha_0 - \alpha_1)*(\alpha_0 + \alpha_1)/2]$$
$$\Delta E_C = [m*(D*\omega_2)^2]*\alpha*\Delta\alpha \tag{81}$$

where $\alpha_0 - \alpha_1 = \Delta\alpha > 0$, the *reduction* in coning angle due to damping,

and $(\alpha_0 + \alpha_1)/2 = \alpha$, the average coning angle over this half cycle.

We now hypothesize that, at least in stable coning flight in which no nutation needs damping, and for *dynamically stable* bullets, the average loss in "forward motion" kinetic energy $\Delta E_\alpha$ over any half coning cycle due to flying with an aerodynamic angle-of-attack $\alpha$ causes the average "frictional damping" decrease in coning energy $\Delta E_C$ during that same half coning cycle. So, these changes in energy must be proportional to each other. That is to say, we are tentatively assuming that a small portion **e** of the yaw-drag of the bullet directly causes the damping of its coning angle $\alpha$ in **steady-state, minimum coning angle, hyper-stable flight**.

In accordance with this hypothesis, we can set $\Delta E_C = e*\Delta E_\alpha$ over any particular half coning cycle, where the constant fraction **e** is greater than **zero** but not greater than **1.0**, and so that

$[m*(D*\omega_2)^2]*\alpha*\Delta\alpha = q*S*V*\alpha^2*e*CD_\alpha*T_2/2$ (82)

or, dividing through by $\alpha^2$ and by $[m*(D*\omega_2)^2]$,

$(\Delta\alpha)/\alpha = T_2*[q*S*V*e*CD_\alpha]/[2*m*(D*\omega_2)^2]$ (83)

We recognize this expression having the form of the classic aeroballistic exponential damping of the coning angle $\alpha$ in damped harmonic motion:

$\alpha(t) = \alpha(0)*\exp[-\lambda_2*t]$ (84)

with

$\lambda_2 = [q*S*V*e*CD_\alpha]/[2*m*(D*\omega_2)^2]$. (85)

Here, $\lambda_2 = -\lambda_S*V/d$ in classic aeroballistic terms.

If we replace the half coning period $T_2/2$ with a small increment in time **dt**, and replace $\Delta\alpha$ per half coning cycle with a small decrease $-d\alpha$ in $\alpha$, then in the limit as **dt** approaches zero, this expression becomes

$d\alpha/\alpha = -\lambda_2*dt$ (86)

After integrating both sides from **0** to **t**,

$\ln[\alpha(t)/\alpha(0)] = -\lambda_2*t$ (87)

Or, after exponentiating

$\alpha(t) = \alpha(0)*\exp[-\lambda_2*t]$ (88)

Here, we have shown the feasibility of arguing that some portion **e** the kinetic energy of the bullet lost due to the yaw-drag attributable to the coning angle-of-attack $\alpha$ during one half of a coning cycle $T_2/2$ matches the reduction in the orbital potential energy (and also in the orbital kinetic energy) of the coning bullet necessary for exponential frictional damping of that coning angle $\alpha$ as a damped harmonic oscillation. **Eq. 85** also gives us a direct formulation of the slow-mode damping factor $\lambda_2$.

So, yaw-drag causes frictional damping which reduces the coning angle. On the other hand, the increased coning angle needed for the coning axis direction to track a step-change in apparent wind direction with response time $T_2/2$ also causes some extra yaw-drag due to that larger coning angle-of-attack. Each coning related mechanism extracts extra kinetic energy from the flying bullet. The rifleman can therefore maximize the kinetic energy delivered by his bullet at long range by minimizing the coning motion of that bullet. Except for damping the coning and nutation of the bullet, the remaining frictional energy loss in flight is dissipated as heat energy.

We coin the term, "hyper-stable flight" for the flight mode in flat-firing in which the coning angle $\alpha$ is reduced to a minimum possible amplitude caused by the downward arcing of the trajectory due to gravity during each half coning cycle $T_2/2$ while also being continually damped. The coning angle of attack $\alpha$ should be less than **0.10-degrees** in hyper-stable flight in flat firing. Thus, with $\delta^2 < 0.000003$, there is essentially no extra yaw-drag and the bullet is flying with its minimum possible total aerodynamic drag force **F_D** attributable only to its zero-yaw drag coefficient **CD_0** at each airspeed.

Importantly, the coning motion of the bullet during this "minimum coning angle" hyper-stable flight mode still allows the rotational cancellation of the aerodynamic lift force **F_L** acting on the bullet due to any crosswinds. Windage corrections would have to be at least an order of magnitude greater if this were not the case. In hyper-stable flight, cross-track windage corrections remain attributable only to a small (**W/V**) cross-track component of the total aerodynamic drag force **F_D** as first formulated by Dedion in 1859, and this drag and its resulting windage are reduced to their minimum possible values for each type of bullet.

Minimum coning angle, hyper-stable flight is achieved earliest in the bullet's flight when the bullet is perfectly launched with zero initial yaw and yaw-rate, when the initial spin-rate of the bullet is very high, when the bullet design is easier to stabilize gyroscopically so that the initial gyroscopic stability of the bullet **Sg ≥ 2.5**, when crosswinds are light and steady, when the density of the ambient atmosphere $\rho$ is relatively low, and when bullet initial velocity **V(0)** is very high. When each of these conditions is met in firing, the perfectly made rifle bullet can achieve hyper-stable flight mode from the very beginning of its ballistic flight where the dynamic pressure $q = (\rho/2)*V^2$ acting to retard that bullet is always its greatest. These hyper-stable rifle bullets should be exceedingly stable while transiting the turbulent transonic speed region far downrange and then continue flying with minimum yaw (coning angle) as subsonic bullets.

Now, let us examine the interchange of kinetic energy in steady-state hyper-stable flight where the coning angle $\alpha < 0.10$ **degrees**. The kinetic energy of forward motion which is lost per half coning cycle due to yaw-drag is given by **Eq. 78** as

$$\Delta E_\alpha(T_2/2) = q*S*V*CD_\alpha*(T_2/2)*\alpha^2 \qquad (78)$$

The kinetic energy of the coning motion is given by **Eq. 80** as

$$E_C = (m/2)*(D*\omega_2)^2 *\alpha^2 \tag{80}$$

We can find the sensitivities of these two expressions to incremental change in coning angle $\alpha$ by finding their partial derivatives with respect to that coning angle $\alpha$

$$\partial(\Delta E)/\partial\alpha = 2*\alpha*q*S*V*CD_\alpha*T_2/2$$

or

$$\partial(\Delta E)/\partial\alpha = \alpha*2\pi*q*S*V*CD_\alpha/\omega_2 \tag{89}$$

and

$$\partial(E_C)/\partial\alpha = m*(D*\omega_2)^2 *\alpha. \tag{90}$$

We now reason that the kinetic energy $\Delta E_C$ of the coning motion lost to frictional damping of the coning angle by the difference $\Delta\alpha$ during each of these "steady-state" half coning cycles must be a small fraction **e** of the kinetic energy $\Delta E$ extracted from the forward motion of the bullet due to that same coning angle difference $\Delta\alpha$ during that same half coning cycle.

For $(\Delta\alpha, \alpha) \neq 0$, we can write

$$e*\Delta\alpha*\partial(\Delta E)/\partial\alpha = \Delta\alpha*\partial(E_C)/\partial\alpha$$

or

$$e*(2\pi*q*S*V/\omega_2)*CD\alpha = m*(D*\omega_2)^2 \tag{91}$$

After substituting for **D** its expression from **Eq. 13** and for $\omega_2$ its expression from **Eq. 22**, and simplifying, this expression reduces to

$$(\omega/2\pi)*d/V = e*k_x^{-2} *CM\alpha*CD\alpha/(CL\alpha + CD_0)^2 \tag{92}$$

In particular, for achieving hyper-stable flight right out of the muzzle where airspeed, dynamic pressure, and drag are always greatest, we can evaluate this relationship using *initial* values:

$$[\omega(0)/2\pi]*d/V_0 = 1/n = e*k_x^{-2} *CM\alpha*CD\alpha/(CL\alpha + CD_0)^2 \tag{93}$$

where  n = Rifling Twist-Rate in calibers/turn

and the aeroballistic coefficients have their values at muzzle airspeed.

For the well studied 30-caliber 168-grain Sierra International bullet launched at Mach 2.5,

L = 3.98 calibers

$k_x^{-2}$ = 9.218 calibers$^{-2}$

$CL\alpha$ = 2.850

$$CD_0 = 0.320$$
$$CM\alpha = 2.560$$
$$CD\alpha = 4.400$$

Greenhill's formula suggests a barrel twist-rate **n** for this bullet of **150/3.98 = 37.7 calibers**.

Substituting and solving for **e**, we find that in this case

**e = 0.002567** (94)

This is just one data point for a rifle bullet which was not even hyper-stable at the muzzle. We need to examine much more data to discover how the energy transfer efficiency ratio **e** behaves for modern rifle bullets in various firing conditions.

However, **Eq. 97**, given the right values, does formulate the rifling twist-rate **n** required to achieve hyper-stable flight as early as possible for modern monolithic copper-alloy ULD bullets which can withstand the required initial spin-rates. Fortunately, the function of the radius of gyration about the spin-axis of the bullet $kx^{-2} = 9.00$ **calibers**$^{-2}$ for almost any solid monolithic ULD rifle bullet.

If we similarly simplify the expression for $\lambda_2$ in **Eq.85** above, we will find the inverse of that the same combination of aeroballistic coefficients which appears as in **Eq. 93**. For brevity, we might term this combination a *Coefficient of Stability* **CSα**, so that at each airspeed

**CSα = (CLα + CD₀)²/(CMα*CDα)** (95)

Coning Theory opens new fields for analytical aeroballistics. We need to exploit it first for practical applications in improving long-range riflery and then determine how much of this work can be generalized to include artillery, mortar, and rocket ballistics.

## Conclusions

We have shown that the CG of the spinning bullet cones around at its gyroscopic precession rate $\omega_2$ in accordance with the Tri-Cyclic Theory, but with its nose angled *inward*, toward its mean trajectory. The magnitude of the rotating coning force {Fc} driving this coning motion of the Bullet's CG is given by:

$$\{F_c\} = q*S*\sin(\alpha)*[CL\alpha+CD] \tag{11}$$

The gyroscopic precession rate $\omega_2$ as found from Tri-Cyclic Theory matches the coning rate $\omega_2$ calculated here as:

$$\omega_2 = q*S*d*CM\alpha/L \tag{22}$$

The CG of the coning bullet maintains a coning distance **D** behind the apex of its coning motion given by the fundamental coning relationship:

$$D = q*S*(CL\alpha+CD)/(m*\omega_2^2) \tag{13}$$

We may sometimes find it more useful to express **D** in calibers **D/d** equivalently as

$$D/d = kx^2 *(Iy/Ix)*(R+1)*(CL\alpha+CD)/CM\alpha \tag{96}$$

We posit that the incoming direction of the apparent wind seen by the moving bullet *continually defines* the orientation of the axis of this coning motion so that the spin-axis of the coning bullet always precesses around the instantaneous eye of the apparent wind. Every wind axes plot shows this.

Furthermore, we posit that the flying bullet can respond to each change in the approaching apparent wind only by *increasing* its coning angle $\alpha$ and its corresponding coning radius **r**. No change in flight conditions could be encountered by the coning bullet which would directly reduce its coning angle $\alpha$.

We found that the coning angle $\alpha(t)$ is itself a *free variable* in Coning Theory which cannot easily be calculated except in special cases. Because *increasing* $\alpha(t)$ is the mechanism by which the coning bullet incorporates changes in flight conditions into its coning motion, the value of $\alpha(t)$ depends upon the complete history of such changes occurring before time **t**.

We theorize that the spin-stabilized bullet in free flight acts exactly as would a similar gimbal-mounted gyroscope under application of similar external torques and torque impulses. We use the gyroscopic terms "precession" and "nutation" in discussing the motions of the bullet's spin-axis direction.

We have shown that the coning motion of the spin-axis of the bullet is caused by a pseudo-regular gyroscopic precession and that the CG of the bullet orbits in a circular helix around its mean trajectory exactly 180 degrees out of phase with the precession of its spin-axis.

We found that higher-rate gyroscopic nutation does not produce significant bullet coning motion, at least not for the example rifle bullets studied here.

The observed vertical deflection of rifle bullets fired through purely horizontal crosswinds has been shown numerically to be a one-time transient effect occurring during initiation of the coning motion upon first encountering that crosswind. We have also investigated the horizontal component of this initial crosswind aerodynamic jump and found that it cannot affect long-range point of impact. The vertical angular effects can be calculated accurately from the measured crosswind at the firing point and incorporated directly into the rifle's aiming system.

We have shown using coning theory that the observed gradual increase of bullet's yaw-of-repose and its consequent horizontal spin-drift are caused by the persistent downward changing due to gravity of the flight path angle of the bullet's trajectory throughout its flight.

Mathematically, we can describe the early coning motion of the spin-axis direction of a rifle bullet upon first encountering a crosswind using complex notation as the vector sum:

$$\alpha(t) = \gamma_0 * i + (\gamma_0 + \delta_0) * \exp[i * \xi_2(t)] + (\delta_0) * \exp[i * \xi_1(t)] \qquad (26)$$

With ongoing time, the slow-mode and fast-mode arms are exponentially damped at their respective aeroballistic damping rates in accordance with linear aeroballistics theory. The eccentric epicyclic sum of these three vectors plotted over ongoing time is shown in the traditional wind axes plots.

We have shown that a small fraction **e** of the dynamically stable bullet's loss of kinetic energy due to yaw-drag is utilized for the exponential damping of the coning angle **α**. We sample evaluated this energy conversion efficiency **e** as **0.002567** for the old Sierra International bullet. In particular, the barrel twist-rate **n** needed to achieve hyper-stable flight right out of the muzzle can now be formulated as

$$1/n = e * k_x^{-2} * CM\alpha * CD\alpha / (CL\alpha + CD_0)^2 = e * k_x^{-2} / CS\alpha \qquad (93)$$

where **n** is the required rifling twist-rate in **calibers per turn**. The required helix angle of the rifling twist is then **180/n** in degrees or **π/n** in radians.

We expect to continue exploiting these energy considerations utilizing Coning Theory as time permits.

## Summary

This new analysis of the basic physics behind the equations of motion describing the flight of a spin-stabilized rifle bullet is entirely consistent with BRL's standard calculus-based formulation of these equations and with the equations of motion currently implemented in existing 6-DoF flight simulation software.

The major accomplishment of this effort is to illuminate the nature of the bullet's coning motion—the circular orbiting of the CG of the bullet about its mean trajectory synchronized with the gyroscopic precession of its spin-axis. The mean CG of the bullet moves smoothly along the mean trajectory with its mean velocity vector tangent to that mean trajectory. The apex of the coning motion points just slightly above, and to the right (for right-hand spinning bullets) of the apparent wind direction approaching the flying bullet. These very small attitude differences are, respectively, the vertical and horizontal dynamic tracking error angles of the coning motion. These tracking errors are quite small in flat firing, but they can become quite large in high-angle artillery and mortar firing. When firing in a wind-free atmosphere, these tracking errors are just above and to the right of the mean velocity vector.

Only the primary aeroballistic forces, drag and lift, and the aeroballistic overturning moment are used in this new analysis of bullet motions. The minor aeroballistic forces and moments (e.g., Magnus force and moment, pitch-damping force and moment, and spin-damping moment, etc.) are not needed to explain analytically each of these observable phenomena of modern supersonic spin-stabilized rifle bullets flying essentially horizontally in flat firing.

Coning theory allows analytic calculation of the one-time vertical aerodynamic jump which occurs whenever a spin-stabilized projectile is fired in a horizontal crosswind. The crosswind aerodynamic jump produces a vertical angular deflection which affects the entire remaining trajectory of the rifle bullet. While this aiming correction is small but still significant in most rifle matches, it would be much larger in firing sideways from a fast-moving aircraft. This transient effect occurs during the first half of the first coning cycle after encountering that purely horizontal crosswind.

We have analyzed and explained numerically the physical causes producing the spinning bullet's yaw-of-repose and its resulting spin-drift phenomena and formulated analytic expressions to calculate them accurately. We discovered that, beyond the first 150 yards or so of flight in flat firing, the horizontal spin-drift of a rifle bullet matches closely an invariant fraction of that bullet's drop from the projected axis of the bore at firing. We developed a formulation to allow an accurate estimation of that fraction.

We discussed certain details of what should, and should not, be seen in the ballistician's traditional wind-axes plots. The yaw-of-repose angle should not appear. The origin of the plot should always represent the direction of the bullet's mean velocity vector at any time during the flight. The epicyclic motions of the bullet's spin-axis direction revolve about the direction of approach of the apparent wind seen by the moving bullet, except for very small horizontal and vertical direction tracking error angular offsets. That is, the coning axis points toward the apparent wind except for these dynamic tracking error angles. The dynamic response of the coning motion to these tracking error angles has a time-constant of one half of the period of the coning motion.

We have shown that the energy needed to realign the coning axis into the approaching apparent wind extracts extra kinetic energy from the flying bullet via increased yaw-drag during one half of a coning cycle. We have also shown that the frictional energy used in damping the coning angle in damped harmonic oscillation is similarly extracted from the kinetic energy of the bullet. We coin the term "hyper-stable flight" for the minimization of these in-flight kinetic energy losses. We explain how riflemen can achieve this steady-state hyper-stable flight from the beginning of ballistic flight of their fired bullets, right out of the muzzle blast cloud.

In short, progress in aeroballistics needs no longer remain stymied by a complete misunderstanding how bullets actually fly through the ambient atmosphere. Practical riflery stands to benefit in many important ways from further exploitation of coning theory.